\documentclass[11pt,letterpaper]{article}
\usepackage{jheppub}
\usepackage{amsfonts,amsmath,latexsym,amssymb,txfonts} 
\usepackage[american]{babel}

\title{Effective Action and Phase Transitions in Thermal Yang-Mills 
Theory on Spheres}

\author{Ivan G. Avramidi and}
\author{Samuel Collopy}
\affiliation{Department of Mathematics,
New Mexico Institute of Mining and Technology,\\
Socorro, NM 87801, USA}
\emailAdd{iavramid@nmt.edu, samuel.collopy@gmail.com}

\abstract{
We study the covariantly constant Savvidy-type chromomagnetic vacuum in finite-temperature Yang-Mills theory on the four-dimen\-sional curved spacetime.
Motivated by the fact that a positive spatial curvature acts as an effective gluon mass we consider the compact Euclidean spacetime $S^1\times S^1\times S^2$, with the radius of the first circle determined by the temperature 
$a_1=(2\pi T)^{-1}$. We show that covariantly constant Yang-Mills fields on $S^2$ cannot be arbitrary but are rather a collection of monopole-antimonopole pairs.
We compute the heat kernels  of all relevant operators exactly
and show that the gluon operator on such a background has negative modes 
for any compact semi-simple gauge group. We compute the infrared regularized effective action and apply the result for the computation of the entropy and the heat capacity of the quark-gluon gas. We compute the heat capacity for the gauge group $SU(2N)$ for a field configuration of $N$ monopole-antimonopole pairs. 
We show that in the high-temperature limit the heat capacity is well defined in 
the infrared limit and exhibits a typical behavior of second-order phase transition $\sim (T-T_c)^{-3/2}$ with the critical temperature 
$T_c=(2\pi a)^{-1}$, where $a$ is the radius of the $2$-sphere $S^2$. 
}


\arxivnumber{ }

\def\II{{\mathbb I}} 
\def\RR{{\mathbb R}} 
\def\CC{{\mathbb C}} 
 
\def\ZZ{{\mathbb Z}}

\def\gfrak{{\mathfrak{g}}}

\def\spinfrak{{\mathfrak{spin}}}

\def\g{\gamma}

\def\m{\mu}

\def\h#1{{\cal #1}}

\def\g{\gamma}

\def\m{\mu}

\def\na{\nabla}

\def\na{\nabla} 
\def\tr{\mathrm{ tr\,}} 
\def\Tr{\mathrm{ Tr\,}}

\def\Det{\mathrm{ Det\,}}

\def\vol{\mathrm{ vol\,}}

\def\const{\mathrm{ const\,}} 

\def\diag{\mathrm{diag\,}} 
\def\Spin{\mathrm{Spin}} 
\def\End{\mathrm{End}} 
\def\Aut{\mathrm{Aut}}

\def\be{\begin{equation}} 
\def\ee{\end{equation}} 
\def\bea{\begin{eqnarray}} 
\def\eea{\end{eqnarray}} 
\def\bed{\begin{definition}{\ }}
\def\eed{\end{definition}}
\def\bd{\begin{description}}
\def\ed{\end{description}}
\def\bc{\begin{center}}
\def\ec{\end{center}}

\newtheorem{definition}{Definition}

\def\sideremark#1{\ifvmode\leavevmode\fi\vadjust
{\vbox to0pt{\vss\hbox to 0pt{\hskip\hsize\hskip1em
\vbox{\hsize2cm\raggedright\pretolerance10000
\noindent{\sf #1}\hfill}\hss}\vbox to8pt{\vfil}\vss}}}
\begin{document}

\maketitle

%=================================================================

\section{Introduction}
\setcounter{equation}0

Despite the tremendous success of quantum chromodynamics (QCD) in
describing the phenomenology of strong interactions of elementary
particles at high energies, a deep understanding of the physics at low
energies is still lacking. At high energies, non-Abelian gauge
theory is asymptotically free, and as a result, perturbation theory
is an adequate tool. However, at low energies, the interaction becomes strong, 
and perturbation theory fails. It has been suggested that this failure is directly linked to
the phenomenon of confinement in QCD, which is a well-known
experimental fact. However, %Despite the overwhelming evidence of confinement, 
the precise nature of a non-pertubative
mechanism ensuring confinement is still not well understood. In
field-theoretic terms, this means that the vacuum of QCD at low energies
has a far more complicated structure than the trivial perturbative one.

A model of a non-perturbative vacuum for an $SU(2)$ gauge theory 
was put forward in 1977 by Savvidy 
\cite{savvidy77}. He proposed an explicit ansatz for the vacuum
gauge fields in form of a constant chromomagnetic field, or more precisely,
a gauge field with covariantly constant field strength in flat four-dimensional Minkowski
space-time with only one nonvanishing color component. 
Savvidy showed that, due   to quantum fluctuations of the gauge
fields, the energy of such a field   configuration is below the
perturbative vacuum level, which leads to infrared instability   of
the perturbative vacuum under creation of a constant chromomagnetic
field. Further   investigations \cite{nielsen78,nielsen79} 
showed that the Savvidy vacuum itself is unstable too,   meaning that
the physical nonperturbative vacuum has an even more complicated
structure. It has been suggested that the real vacuum is likely to have
a small domain  structure with random constant chromomagnetic fields
(spaghetti vacuum).

In our papers \cite{avramidi95a,avramidi99} we extended Savvidy's
investigation  by considering more complicated gauge groups and flat
spacetimes of dimension higher than four. We showed that for an arbitrary
compact simple gauge group in dimensions higher than four 
there exist more general nontrivial field
configurations with several color and space-time components that turn
out to be stable. In \cite{avramidi99} we proposed an explicit example
of such   background field configurations. 

In the present paper we propose a new mechanism to stabilize the Savvidy
vacuum in four dimensions. The main idea of this approach is that a 
{\it positive space curvature could  provide an effective mass term} for
the gauge fields on the chromomagnetic vacuum, thus, making the vacuum
stable. To simplify the calculations, we consider the
space-times with compact space slices with 
the product structure $S^1\times S^2$. 
To also study the finite-temperature 
effects we consider Euclidean spacetimes 
of the form $S^1\times S^1\times S^2$. 

However, as we show below,  
topological considerations on the sphere constrain the magnetic field to be
of the same order of magnitude as the space curvature, thus negating the
stabilization effect of the curvature term. Moreover, even under these
constraints, an interesting second-order phase transition occurs at a 
critical temperature near the inverse radius of the sphere.

%=======================================================================
\section{Yang-Mills Theory}
\setcounter{equation}0

Let $(M,g)$ be an $n$-dimensional
pseudo-Riemannian  orientable spin manifold without boundary with
a globally hyperbolic
metric $g$. 
We denote the local coordinates on $M$ by $x^\mu$, with Greek indices
running over $0, 1,\dots, n-1$. 
We denote the frame indices by the low case Latin indices
from the beginning of the alphabet, which
also run over $0,1,\dots,n-1$.
The frame indices should not be confused with the group indices introduced below
that are enclosed in parenthesis. We use Einstein summation convention and sum over repeated indices.
The coordinate indices are raised and lowered by the metric tensor
$g_{\mu\nu}$ and the frame indices are raised and lowered by the Minkowski
metric, $\eta_{ab}$.

We choose a local Lorentz frame on the tangent bundle $TM$,
$e_a=e_a{}^\mu \partial_\mu$, and 
the dual frame on the cotangent bundle $T^*M$,
$\sigma^a=\sigma^a{}_\mu\,dx^\mu$.
We denote the corresponding spin connection $1$-form by
$\omega^{a}{}_b=\omega^{a}{}_{b\mu}\,dx^\mu$ and
its curvature $2$-form by
$\Theta^{a}{}_b=\frac{1}{2}\Theta^{a}{}_{b\mu\nu}\,dx^\mu\wedge dx^\nu$ so that
$R_{\alpha\nu}= e_{a}{}^\mu e^{b}{}_{\alpha} \Theta^{a}{}_{b\mu\nu}$ is
the Ricci tensor, and 
$R=g^{\mu\nu}R_{\mu\nu}$
is the scalar curvature.
Let $\mathcal{T}$ be a spin-tensor
bundle realizing a representation $T$ of the spin group,
\mbox{$\Spin(1,n-1)$}, with generators $\Sigma_{ab}$.
The spin connection induces a connection $\nabla^T$
on the bundle ${\cal T}$ 
with the curvature
$
\mathcal{R}_{\mu\nu}=\frac{1}{2}\Theta ^{ab}{}_{\mu\nu}
T(\Sigma_{ab})\,.
$
Recall that the generators of the spin group 
in the vector representation
and the spinor
representation are
\be
T_1(\Sigma_{ab}){}^c{}_d=2\delta^{c}{}_{[a}\eta_{b]d}\,,
\qquad
T_{\rm spin}(\Sigma_{ab})
=\frac{1}{2}\gamma_{ab}\,,
\ee
where $\eta_{bd}$ is the Minkowski metric and
$\gamma_{ab}=\gamma_{[a}\gamma_{b]}$.
Here and everywhere below the square brackets denote the antisymmetrization
over indices included.

Let $G$ be a $m$-dimensional compact simple Lie group and
$\gfrak$ be its Lie algebra. We use lower case gothic letters
to denote Lie algebras, for example, $\spinfrak(1,n-1)$.
We denote the group indices, which run over $1,2,\dots,m$,, 
by the low case Latin letters from the middle of the alphabet.
Let $C^i{}_{jk}$ be the structure constants of $G$ in a given basis.
The  real $m\times m$ matrices $C_{i}$ defined by $(C_{i})^{j}{}_{k}
=C^{j}{}_{ik}$
form a basis in the Lie algebra $\gfrak$ and define the adjoint
representation ${\rm ad}: \gfrak\to \End(\RR^m)$ of the algebra $\gfrak$
by endomorphisms of the vector space $\RR^m$. Of course, this also defines the adjoint representation ${\rm Ad}: G\to \Aut(\RR^m)$ of the group $G$
into the automorphism group of 
the vector space $\RR^m$. In the following we will identify the 
algebra $\gfrak$ with its adjoint representation.
The Cartan-Killing metric $\gamma_{ij}$ on the Lie algebra
$\gfrak$ is defined by
\be
\tr C_iC_j=C^{k}{}_{il}C^{l}{}_{jk} = -2\gamma_{ij}\,,
\label{22xx}
\ee
the precise form depends, of course, on the choice of the basis. We will
determine it later.
For compact semi-simple groups it is positive definite.
We will use it
to raise and lower the group indices.

We consider  the principal fiber
bundle over the manifold $M$  with 
the structure group $G$ and
the typical fiber $G$.
Let $\rho_W: G\to {\rm Aut}(W)$ be an irreducible 
representation of the group $G$ into the automorphism group of 
an $N$-dimensional (real or complex) vector space $W$.
Sometimes, we will denote the representation $\rho_W$ by the vector space
$W$; this should not cause any confusion. For example, we will denote
the generators of this representation by $W(C_{i})$.
Let  ${\cal W}$ be the associated vector bundle with the structure
group $G$ and the typical fiber $W$.
Then for any
spin-tensor bundle ${\cal T}$ (realizing a representation $T$ of the 
spin group, ${\rm Spin}(1,n-1)$) the vector bundle $\mathcal{V}={\cal
W}\otimes {\cal T}$ is a twisted spin-tensor bundle
realizing the representation
$
V=W\otimes T\,.
$
The sections of the
bundle ${\cal V}$ are represented locally by 
(real or complex) $N$-tuples of spin-tensors.

Let ${\cal A}_\mu=A^{i}{}_\mu C_{i}$ and 
${\cal A}={\cal A}_\mu dx^\mu$ 
be the Yang-Mills connection $1$-form
taking values in the Lie algebra $\gfrak$
and 
${\cal F}_{\mu\nu}={\cal F}^i{}_{\mu\nu}C_i
$
and ${\cal F}=\frac{1}{2}{\cal F}_{\mu\nu}\,dx^\mu\wedge dx^\nu
=d{\cal A}+{\cal A}\wedge{\cal A}$ be its curvature 2-form.
It is worth remembering that in a non-trivial bundle, there are several 
overlapping coordinate patches; the connection one-form is not globally well-defined,
in general, but is rather described by a collection of its representations in each patch,
which are related in overlapping patches
by gauge transformations.

Let $\nabla^V$ be the total connection on the twisted spin-tensor bundle
${\cal V}$.
The curvature of the total connection
on the twisted spin-tensor bundle is
equal to
$
\II_W\otimes T({\cal R}_{\mu\nu})
+W({\cal F}_{\mu\nu})\otimes\II_T
\,.
$
We will usually omit the identity matrices for the sake of simplicity
of notation.
The covariant Laplacian $\Delta_{T\otimes W}=g^{\mu\nu}\nabla^V_\mu\nabla^V_\nu$
acting on sections of the twisted spin-tensor bundle ${\cal V}$ has the form
\be
\Delta_{T\otimes W}=
g^{-1/2}\left[\partial_\mu+\frac{1}{2}\omega^{ab}{}_\mu T(\Sigma_{ab})
+W({\cal A}_\mu)\right]
g^{1/2}g^{\mu\nu}
\left[\partial_\nu+\frac{1}{2}\omega^{cd}{}_\nu T(\Sigma_{cd})+W({\cal A}_\nu)\right]\,.
\ee

We will use twisted Lie derivatives defined as follows.
Suppose that there is a faithful representation
$\rho_X: {\rm Spin}(1,n-1)\to {\rm Aut}(W)$ of the 
spin group in the same vector space $W$ with generators $X(\Sigma_{ab})$.
Then the matrices
\be
G_{ab}=\II_W\otimes T(\Sigma_{ab})-X(\Sigma_{ab})\otimes\II_T\,
\ee
are the generators of the twisted representation 
$\rho_{W\otimes T}: {\rm Spin}(1,n-1)\to {\rm Aut}(V)$ 
of the spin group.
Let $\xi$ be a Killing vector field.
The twisted
Lie derivative of sections of the vector bundle ${\cal V}$
along $\xi$ is defined by
\bea
{\cal L}_\xi
&=&\xi^\mu\nabla^V_\mu-\frac{1}{2}\xi_{[a;b]}G^{ab}
\nonumber\\
&=&\xi^\mu\partial_\mu
+\frac{1}{2}\left[\xi^\mu\omega_{ab}{}_\mu 
-\xi_{[a;b]}
\right]T(\Sigma^{ab})
+\xi^\mu A^{i}{}_\mu
W(C_{i})
+\frac{1}{2}\xi_{[a;b]}X(\Sigma^{ab})
\,.
\eea

The action of the Yang-Mills theory in curved spacetime is constructed as follows.
We consider two associated vector bundles ${\cal W}_{\rm spin}$
and ${\cal W}_0$ of dimensions $N_{\rm spin}$ and $N_0$ respectively
realizing some irreducible
representations, $W_{\rm spin}$ and $W_0$, of the gauge group.
Usually, the representation $W_{\rm spin}$ realized by the spinor fields
is taken to be the fundamental (or defining) representation of the
gauge group.
The scalar fields are just sections of the bundle ${\cal W}_0$,
whereas the spinor fields are sections of the spinor bundle twisted by 
${\cal W}_{\rm spin}$.
Then the  
classical action of the model is the functional
\bea
S &=& -\int\limits_M
dx\; g^{1/2}\Biggl\{
%\frac{1}{f^2}\left(R_{\mu\nu}R^{\mu\nu}-\frac{1}{3}R^2\right)
%-\frac{1}{6\nu^2}R^2
%-\frac{1}{2\varkappa}(R-2\Lambda)
%\nonumber\\
%&&
%+
\frac{1}{2e^2}|{\cal F}|^2
+\left<\psi,[\g^\m\na_\m+M]\psi\right>_{W_{\rm spin}} 
+\frac{1}{2}\left<\nabla^\mu\varphi,\nabla_\mu\varphi\right>_{W_0}
+V(\varphi)\Biggr\}\,,
\nonumber\\
\eea
where 
$g=\det g_{\mu\nu}$, 
$|{\cal F}|^2
=-\frac{1}{4}\tr_{Ad}{\cal F}_{\mu\nu}{\cal F}^{\mu\nu}$,
%$f^2$ is the Weyl coupling constant, 
%$\nu^2$ is the conformal coupling constant,
%$\varkappa=8\pi G$ is the Einstein coupling constant,
$e$ is the Yang-Mills coupling constant,  $V(\varphi)$ is a
potential for scalar fields (such that $V(0)=V'(0)=0$) 
and  $M$ is the 
spinor mass matrix.
%Here we included the terms quadratic in curvature to make the theory
%renormalizable.

%===================================================================
\section{Effective Action}
\setcounter{equation}0

In this paper we will 
not quantize gravity, assuming the gravitational field to be classical and simply ignoring 
all quantum-gravitational effects. The energy scale of our primary interest will be well below
the Planckian scale, so that this assumption is reasonable in any theory of quantum gravity.
  
The effective action is expressed in terms of functional
determinants of differential operators
(see, for example, \cite{dewitt65,avramidi10b}). 
Such determinants can be
defined, strictly
speaking, only for elliptic operators on compact manifolds
by making use of a  regularization procedure, for
example, zeta-function regularization. For hyperbolic operators and
for non-compact manifolds  such determinants do not have a direct
mathematical meaning.

That is why we will make the further assumption that the background
is static, that is, there is a global time-like Killing vector field
$\partial_t$. Moreover, we will assume that
the spacetime has the simple structure of a product manifold
$M=\RR\times \Sigma$ and
that all background fields are static
and do not have time-like components.

In the case of static background
one can make an analytic continuation to a purely
imaginary time 
$
t\to i\tau\,,
$
with a positive-definite Riemannian metric.  
Moreover, we can go even further and compactify the Euclidean time
by replacing $\RR$ by a circle $S^1$ of radius $a_1$, that is, by restricting the range,
$
0\le \tau\le\beta\,,
$
where $\beta=2\pi a_1$ is the circumference of the circle $S^1$ and requiring all fields
to be periodic in the Euclidean time $\tau$ with period $\beta$.  The
``Euclidean'' space-time $M=S^1\times \Sigma$ is then a compact
manifold and Lorentzian spin group ${\rm Spin}(1,n-1)$ becomes the 
Euclidean spin group
${\rm Spin}(n)$, which is compact. 
This corresponds to a statistical ensemble at a finite
temperature $T=1/\beta$. In the limit of infinite
radius $\beta\to \infty$ we recover the
zero-temperature theory.

We will consider a background in which there are no matter fields
and Yang-Mills fields and gravitational field are covariantly constant (parallel), that is,
\be
\nabla_\mu R_{\rho\sigma\alpha\beta}=0\,,\qquad
\nabla_\mu {\cal F}_{\alpha\beta}=0\,.
\ee
More precisely, we will study the case when $\Sigma=S^1\times S^2$, that is, $M=S^1\times S^1\times S^2$.

There exists a minimal gauge such that all
differential operators involved are second-order differential operators
of Laplace type
\be
L=-\Delta+Q
\ee
with some endomorphism $Q$. 
The precise nature of the Laplacian
$\Delta=g^{\mu\nu}\nabla_\mu\nabla_\nu$
depends, of course,
on the fields it is acting upon. 

Let $L^{YM}_{\rm ghost}$ be the Laplacian acting on scalar fields in 
adjoint representation of the gauge group
defined by
\be
L^{YM}_{\rm ghost}=-\Delta_{T_0\otimes Ad}\,.
\ee
Let $Q_{\rm vect}$ be the endomorphism acting on vectors in adjoint representation
of the gauge group defined by
\be
(Q_{\rm vect}\varphi)^a=\left(R^a{}_b\II_{Ad}-2{\cal F}^a{}_b\right)\varphi^b\,.
\ee
Let $L_{\rm vect}$ be  an operator acting on vector fields in adjoint
representation of the gauge group defined by
\be
L_{\rm vect}=-\Delta_{T_1\otimes Ad}+Q_{\rm vect}\,.
\ee

We will suppose that 
the spinor mass matrix $M$ commutes with the Dirac operator
$\gamma^\mu\nabla_\mu$.
Let $Q_{\rm spin}$ be the endomorphism acting on
spinor fields in the representation $W_{\rm spin}$ 
of the gauge group defined by
\be
Q_{\rm spin}=
\frac{1}{4}R\II_{T_{\rm spin}}\otimes \II_{W_{\rm spin}}
-\frac{1}{2}\g^{ab}W_{\rm spin}({\h F}_{ab}) 
+ M^2\,.
\ee
Let $L_{\rm spin}$ be the differential operator acting on
spinor fields in the representation $W_{\rm spin}$ 
of the gauge group defined by
\be
L_{\rm spin} = -\Delta_{T_{\rm spin}\otimes W_{\rm spin}}
+Q_{\rm spin} 
\,.
\ee

Let $L_0$ be a differential operator 
acting on   scalar fields in some 
representation $W_0$ of the gauge group defined by
\be
L_0=-\Delta_{T_0\otimes W_0}+Q_0\,,
\ee
where $Q_0$ is a matrix defined by
\be
\frac{d^2}{d\varepsilon^2}V(\varepsilon h)\Big|_{\varepsilon=0}
=\left<h, Q_0h\right>_{W_0}\,.
\ee

The most important observation that should be made at this point is that the 
positive curvature acts as a mass (or positive potential) term in both the Yang-Mills operator
and the spinor operator. While the magnetic field reduces the eigenvalues of the Yang-Mills operator the positive Ricci tensor increases them. Roughly speaking, {\it it is the balance of these two terms that determines whether or not the Yang-Mills operator is positive} (so that the vacuum is stable).

In the minimal gauge  the one-loop effective action
is given by
\cite{avramidi95a,avramidi99}
\be
\Gamma = S+\hbar\left(\Gamma_{(1)YM} + \Gamma_{(1){\rm mat}}\right)
+O(\hbar^2)\,,
\label{316}
\ee
where 
\be
\Gamma_{(1)YM} = \frac{1}{2}\;\log\Det L_{\rm vect}
-\log\Det L^{YM}_{\rm ghost}\,,
\ee
and
\be
\Gamma_{(1)\rm mat} = 
\frac{1}{2}\;\log\Det L_0
-\frac{1}{2}\;\log\Det L_{\rm spin}
\,,
\ee
are the contributions of the Yang-Mills field,
and matter fields respectively,
and $\Det$ is the functional determinant.
The Planck constant is introduced here just for illustrative purposes.
Henceforth, we set $\hbar=1$.

The operators introduced above are second-order elliptic partial
differential operators on compact manifold.
The spectrum of these operators depends,
of course, on the background fields.
Elliptic operators on compact manifolds can only have a finite number of
negative eigenvalues. The negative eigenvalues indicate instability of the 
vacuum at low energies. 

That is why, to study the infrared behavior of the system, one has to 
introduce an infrared regularization 
and take it off at the very end as in 
\cite{avramidi99}. 
For example, one could introduce a sufficiently large mass parameter $z$
so that all operators are positive, which is equivalent to replacing the operators $L$ by $L+z$, and study the dependence of the 
infrared regularized effective action on $z$.
If there are no infrared divergences then there is a well defined limit $z\to 0$. In the case of non--trivial low-energy behavior there appears an imaginary part of the effective action or some infrared 
logarithmic singularities.
One should stress that 
although the ultraviolet regularization is a rather formal method, the infrared regularization parameter can take, in 
principle, a {\it direct physical meaning}, something like $\Lambda_{QCD}$.

We will assume that this has been done, that is, we will add a mass parameter
to the Yang-Mills and the ghost operators so that all operators are positive.
The determinants of {\it positive}
elliptic operators can be regularized by the zeta-function
regularization method which can be
summarized by
\cite{avramidi91,avramidi00,avramidi10b}
\be
\log\Det(L+z)=-\zeta'(0)\,,
\label{3.54}
\ee
where $\zeta'(0)=\frac{\partial}{\partial s}\zeta(s)\Big|_{s=0}$
and
\be
\zeta(s)=
\sum_{k=1}^\infty d_k \left(
\frac{\lambda_k+z}{\mu^2}\right)^{-s}
\,
\ee
where $\lambda_k=\lambda_k(L)$ 
are the eigenvalues of the operator $L$ and $d_k$ are their multiplicities.
The analytic continuation of the zeta function gives a meromorphic function
of $s$, which is analytic at $s=0$; therefore, 
the determinant (\ref{3.54}) is a well defined invariant.

We can also express the zeta-function in terms of the heat trace
of the operator
$L$ defined by
\be
\Tr\exp(-tL)=
\sum_{k=1}^\infty d_k e^{-t\lambda_k}
=\int_M dx\; g^{1/2}\tr U^{\rm diag}_L(t)
\,,
\ee
where $U_L(t;x,x')=\exp(-tL)\delta(x,x')$ is the heat kernel
of the operator $L$ and
$
U^{\rm diag}_L(t;x)=U_L(t;x,x)\,
$
is the heat kernel diagonal.
Then the zeta-function is related to the heat trace by
the Mellin transform
\be
\zeta(s)=\frac{\mu^{2s}}{\Gamma(s)}
\int\limits_0^\infty dt\;t^{s-1}e^{-tz}\Tr\exp(-tL)\,,
\ee
where $\mu$ is a renormalization parameter introduced to preserve
dimensions.

We introduce a useful function $\Theta_L(t)$ as follows
\be
\Theta_L(t)=(4\pi t)^{n/2}\Tr\exp(-tL)\,.
\ee
It is well-known that as $t\to 0$
\be
\Theta_L(t)\sim \sum_{k=0}^\infty
B_k t^{k}\,,
\ee
where $B_k=B_k(L)$ are some spectral invariants of the operator $L$.
The first three coefficients have the form
\cite{dewitt65,avramidi91,avramidi00}
\bea
B_0&=&\int_{M}dx g^{1/2}\tr \II\,,
\\[10pt]
B_1&=&\int_{M}dx g^{1/2}\tr\left(\frac{1}{6}R\II-Q\right),\qquad
\\[10pt]
B_2&=&\int_{M}dx g^{1/2}\tr\Biggl\{
\frac{1}{2}\left(\frac{1}{6}R\II-Q\right)^2
+\frac{1}{180}R_{abcd}R^{abcd}\II
-\frac{1}{180}R_{ab}R^{ab}\II
\nonumber\\
&&
+\frac{1}{12}\left[{\cal R}_{ab}+{\cal F}_{ab}\right]
\left[{\cal R}^{ab}+{\cal F}^{ab}\right]
%-\frac{1}{6}\Delta Q+\frac{1}{30}\Delta R
\Biggr\}\,.
\label{317}
\eea

It will be convenient to represent the heat trace as $t\to 0$ as follows
\be
\Theta_L(t)\sim e^{-t\lambda}\sum_{k=0}^\infty
A_k(\lambda) t^{k}\,,
\label{322xx}
\ee
where $\lambda$ is a new arbitrary
parameter (that should not be confused with $z$)
and  
\be
A_k(\lambda)=\sum_{j=0}^k\frac{1}{j!}\lambda^{j}B_{k-j}\,.
\ee

The analytic continuation of the zeta function to $s=0$ can be obtained by integration by parts. 
Then the zeta-regularized determinant can be expressed directly in terms of an integral of the heat trace. For even $n$ we obtain
\be
\log\Det_\mu(L+z)=
\frac{(4\pi)^{-n/2}}{\Gamma\left(1+\frac{n}{2}\right)}
\int_0^\infty dt\;\left[\log(\mu^2 t)
+\Psi\left(1+\frac{n}{2}\right)\right]
\left(\frac{\partial}{\partial t}\right)^{1+\frac{n}{2}}
\left[e^{-tz}\Theta_L(t)\right]\,,
\ee
where $\Psi(s)=\Gamma'(s)/\Gamma(s)$ is the logarithmic derivative of the Gamma function.

Let us consider the case of four dimensions, $n=4$, in more detail.
Then
\be
\log\Det_\mu(L+z)=
\frac{1}{2}(4\pi)^{-2}
\int_0^\infty dt\;\left[\log(\mu^2 t)
+\frac{3}{2}-\CC\right]
\left(\frac{\partial}{\partial t}\right)^{3}
\left[e^{-tz}
\Theta_L(t)\right]\,,
\ee
where $\CC\approx 0.58...$ is the Euler constant.
It is not difficult to find the dependence of the determinant on the renormalization parameter $\mu$
\be
\log\Det_\mu(L+z)=-(4\pi)^{-2}\log\frac{\mu^2}{\lambda}\; 
A_2(-z) 
+\log\Det_{\sqrt{\lambda}}(L+z)\,.
\ee
Now, let 
\bea
\Theta^{\rm ren}_L(t)
&=&\Theta_L(t)
-e^{-t\lambda}\left[A_0
+A_1(\lambda)t
+A_2(\lambda)t^{2}\right]\,,
\label{323x}
\eea
The renormalized heat trace has the following asymptotics: as $t\to 0$
\be
\Theta^{\rm ren}_L(t)=A_3(\lambda)t^3+O(t^4)\,,
\ee
and as $t\to \infty$
\be
\Theta^{\rm ren}_L(t)=-e^{-t\lambda}
\left[A_2(\lambda)t^2
+A_1(\lambda)t+A_{0}(\lambda)\right]
+(4\pi)^2 d_1t^2e^{-t\lambda_1}
+O(e^{-t\lambda_2})\,.
\ee
Thus, we can consider the integral
\bea
\log\Det_{\rm ren}(L+z)
&=&-(4\pi)^{-2}\int_0^\infty \frac{dt}{t^3}e^{-tz}\;
\Theta^{\rm ren}_L(t)\,,
\eea
which is well defined since it converges both at $0$ and $\infty$.

One can compute the dependence of the renormalized determinant on $\lambda$
exactly. First, we show that
\be
\lambda\frac{\partial}{\partial\lambda}\log\Det_{\rm ren}(L+z)
=-(4\pi)^{-2}\left(A_2(-z)+\lambda A_1(-z)
+\frac{1}{2}\lambda^2A_0\right)\,.
\ee
Then, by integrating this equation we get
\bea
\log\Det_{\rm ren}(L+z)
&=&-(4\pi)^{-2}\left[A_2(-z)\log\frac{\lambda}{\lambda_0}
+\lambda A_1(-z)
+\frac{1}{4}\lambda^2 A_0\right]
%\nonumber\\
%&&
+\const\,,
\nonumber\\
\eea
where $\lambda_0$ is some constant.
Notice that in the limit
when $\lambda\to 0$ there is
an infrared divergence
\be
\log\Det_{\rm ren}(L+z)=-(4\pi)^{-2}
A_{2}(-z)\log\frac{\lambda}{\lambda_0}
+O(1)
\,.
\label{323}
\ee

Now, by integrating by parts one can show that
\be
\log \Det_\mu(L+z) = \log\Det_{\rm ren}(L+z)
-(4\pi)^{-2}\log\frac{\mu^2}{\lambda}\; A_2(-z)
+c_0\lambda^2 A_0+c_1 \lambda A_1(-z) +c_2 A_2(-z)
\,.
\ee
Here $c_0, c_1$ and $c_2$ are some numerical constants dependent on the regularization scheme, in particular, they can be set to zero without loss of generality.

By using this regularization of functional determinants we obtain the 
effective action in the form
\bea
\Gamma^{}_{(1)YM}&=&
-\frac{1}{2}(4\pi)^{-2}\left\{
\beta_{YM}\log\frac{\mu^2}{\lambda}
+
\int_0^\infty \frac{dt}{t^3}\;e^{-tz}
\Theta^{\rm ren}_{YM}(t)
\right\}
\,,
\label{332}
\\[10pt]
\Gamma^{}_{(1){\rm mat}}&=&
-\frac{1}{2}(4\pi)^{-2}\left\{
\beta_{\rm mat}\log\frac{\mu^2}{\lambda}
+\int_0^\infty \frac{dt}{t^3}\;
\Theta^{\rm ren}_{\rm mat}(t)
\right\}
\,,
\label{332a}
\eea
where 
\bea
\Theta^{\rm ren}_{YM}(t)
&=&\Theta^{\rm ren}_{L_{\rm vect}}(t)
-2\Theta^{\rm ren}_{L^{YM}_{\rm ghost}}(t)\,,
\\[10pt]
\Theta^{\rm ren}_{\rm mat}(t)
&=&\Theta^{\rm ren}_{L_{0}}(t)
-\Theta^{\rm ren}_{L_{\rm spin}}(t)\,,
\eea
and
\bea
\beta_{YM}&=&
B_2(L_{\rm vect})
-zB_1(L_{\rm vect})
+\frac{z^2}{2}B_0(L_{\rm vect})
\nonumber\\
&&
-2B_2(L^{YM}_{\rm ghost})
+2zB_1(L_{\rm ghost})
-z^2B_0(L_{\rm ghost})
\,,
\\[10pt]
\beta_{\rm mat}&=&
B_2(L_{0})
-B_2(L_{\rm spin})\,.
\eea

The main idea of the renormalization group is based on the realization that
the total effective action, $\Gamma=S+\hbar \Gamma_{(1)}+\cdots$, should not depend on the arbitrary
renormalization parameter $\mu$. This means that 
in renormalizable field theories
the coupling constants in the classical action should depend on $\mu$ in such a way to exactly compensate the dependence of the one-loop effective action on $\mu$, that is,
\be
\mu\frac{\partial}{\partial\mu}S=-
\mu\frac{\partial}{\partial\mu}\Gamma_{(1)}
=(4\pi)^{-2}\left(
\beta_{YM}+\beta_{\rm mat}
\right)\,.
\ee
This means that the classical action should have terms of the same type
as those in the coefficients $\beta_{YM}$ and $\beta_{\rm mat}$.
%Since these coefficients contain terms quadratic in the curvature,
%the Einstein general relativity is not renormalizable. 
%To make it renormalizable one would need to include in the classical action %quadratic terms in the curvature, which has its own problems, most importantly, %non-unitarity.

As we discussed above, all relevant operators are of Laplace type
$L=-\Delta+Q$. 
For a covariantly constant background 
the endomorphism $Q$ is covariantly constant, and, therefore, commutes
with the Laplacian. Therefore, the heat semigroup of the operator $L$ is
determined by the heat semigroup of the Laplacian
\be
\exp(-tL)=\exp(-tQ)\exp(t\Delta)\,,
\ee
so, the heat kernel diagonal of the operator $L$ has the form
\be
U^{\rm diag}_L(t)=\exp(-tQ)U^{\rm diag}(t)\,,
\ee
where $U^{\rm diag}(t)$
denotes the heat kernel diagonal of the pure 
Laplacian acting on a vector bundle ${\cal V}$.

By using this property we can express the heat kernel diagonals
of the operators introduced above in terms of the heat kernel
of the corresponding Laplacians,
\bea
U^{\rm diag}_{L_{\rm vect}}(t)&=&\exp\left(-tQ_{\rm vect}\right)
U^{\rm diag}_{T_1\otimes Ad}(t)\,,
\\
U^{\rm diag}_{L^{YM}_{\rm ghost}}(t)&=&U^{\rm diag}_{T_0\otimes Ad}(t)\,,
\\
U^{\rm diag}_{L_0}(t)&=&\exp\left(-tQ_0\right)
U^{\rm diag}_{T_0\otimes W_0}(t)\,,
\\
U^{\rm diag}_{L_{\rm spin}}(t)
&=&\exp\left(-tQ_{\rm spin}\right)
U^{\rm diag}_{T_{\rm spin}\otimes W_{\rm spin}}(t)\,.
\eea
This means that
\bea
\Theta_{YM}(t)
&=&\vol(M)(4\pi t)^2\tr_{Ad}\Big[\tr_{T_1}\exp(-tQ_{\rm vect})
U^{\rm diag}_{T_1\otimes Ad}(t)
-2U^{\rm diag}_{T_0\otimes Ad}(t)\Big]\,,
\\[12pt]
\Theta_{\rm mat}(t)&=&
\vol(M)(4\pi t)^2
\Biggl\{\tr_{W_0}\exp(-tQ_0)U^{\rm diag}_{T_0\otimes W_0}(t)
\nonumber\\
&&-\tr_{W_{\rm spin}}\tr_{T_{\rm spin}}
\exp(-tQ_{\rm spin})
U^{\rm diag}_{T_{\rm spin}\otimes W_{\rm spin}}(t)
\Biggr\}\,.
\eea
The renormalized functions $\Theta^{\rm ren}_{YM}(t)$ and 
$\Theta^{\rm ren}_{\rm mat}(t)$ are obtained from this by finding
the smallest eigenvalue and then subtracting some terms according to
the prescription (\ref{323x}).
Thus, we need
to compute the heat kernel diagonals for the Laplacians only.

%=====================================================
\section{Geometry of the Sphere $S^2$}

In this section we follow mainly our paper
\cite{avramidi09}.

%=========================
\subsection{Metric}

We cover the sphere $S^2$ (of radius $a$)
by two coordinate patches: one patch covering the South pole and another patch covering the North pole.
We will use
the 
spherical coordinates $(r,\varphi)$, which range over $0\le r\le a\pi$ and $0\le \varphi \le 2\pi$.
The South coordinate patch is the neighborhood of the South pole $r=0$,
whereas the North coordinate patch is the neighborhood of the North pole
$r=a\pi$.
The volume of $S^2$ is, of course,
$
\vol(S^2)=4\pi a^2\,.
$

The metric in spherical coordinates is
\be
ds^2=dr^2+a^2\sin^2(r/a)d\varphi^2\,.
\ee
We choose an orthonormal basis of $1$-forms
\bea
\sigma^1&=&\cos\varphi\, dr- a\sin(r/a)\sin\varphi\,d\varphi\,,
\\
\sigma^2&=&\sin\varphi\, dr+ a\sin(r/a)\cos\varphi\,d\varphi\,.
\eea
Then the spin connection one-form is
\bea
\omega_{ab}
&=&
\varepsilon_{ab}[1-\cos(r/a)]\;d\varphi
\,,
\eea
the Riemann curvature is
\be
R_{abcd}=\frac{1}{a^2}\varepsilon_{ab}\varepsilon_{cd}
=\frac{1}{a^2}(\delta_{ac}\delta_{bd}-\delta_{ad}\delta_{bc})\,,
\ee
and the Ricci tensor and the scalar curvature are
\be
R_{ab}=\frac{1}{a^2}\delta_{ab}\,,
\qquad
R=\frac{2}{a^2}\,.
\ee

%=======================
\subsection{Connection}

In two dimensions
the Yang-Mills connection with a covariantly constant curvature
is necessarily Abelian and
can always be chosen to be proportional to the spin connection, that is,
\be
{\cal A}=-X[1-\cos(r/a)]\,d\varphi\,,
\ee
where $X={\rm ad}(\Sigma)$ is an $m\times m$ real anti-symmetric matrix.
The Yang-Mills curvature $2$-form is
\be
{\cal F}
=-X\frac{1}{a}\sin(r/a)\;dr\wedge d\varphi
\,,
\ee
in components, 
$
{\cal F}_{cd}=\varepsilon_{cd}H\,,
$
where
$
H=-X/a^2\,.
$
The connection has to be redefined (via a gauge transformation) in the North coordinate patch. 
By defining
\be
{\cal A}'=X[1+\cos(r/a)]\,d\varphi\,,
\ee
we obtain the connection that is well defined globally.
Now, we have that
\be
{\cal A}'-{\cal A}=dU U^{-1}\,,
\ee
where
$
U=\exp\left(2X\varphi\right)\,
$
is an element of the group $G$. 

Since $U$ should be periodic
in $\varphi$, the matrix $X$ should satisfy the condition
\be
\exp\left(4\pi X\right)=\II\,,
\ee
which means that its eigenvalues must be either zero or imaginary half-integers. 
Only in this case the vector bundle is globally defined.

Since $X$ is anti-symmetric, its non-zero eigenvalues must appear in pairs.
That is, the spectrum of the matrix $X$ must be
\be
{\rm Spec}(X)=\left\{
\underbrace{0,\dots,0}_{r}, i\frac{n_1}{2}, -i\frac{n_1}{2},\dots, 
i\frac{n_p}{2}, -i\frac{n_p}{2}
\right\}\,,
\ee
where $r=m-2p$, and $n_j$, $j=1,\dots,p$, are some non-zero integers.

Such construction for each $n$ is nothing but 
the Hopf complex line bundle ${\cal H}_n$ over $S^2$, which is equal to a tensor product of $n$
Hopf bundles ${\cal H}_1$ (or dual Hopf bundles for $n<0$).
This is sometimes called the topological quantization condition.
The deep fundamental reason for this condition is the Chern theorem
\cite{frankel97}, which in 
particular says that the Chern form
\be
\frac{i}{2\pi}\int_{S^2} {\cal F}=-2iX
\ee
has integer eigenvalues.
Note that the matrix $X$ is the generator of a representation 
$\rho_X: {\rm Spin}(2)\to \Aut(\RR^m)$
of the double cover of the group $SO(2)$
(which we, by definition, still call the spin group ${\rm Spin}(2)$
in two dimensions).

It is important to understand that the numbers $n_j$ are not independent.
Because $X$ lies in the adjoint representation of the compact semi-simple Lie algebra $\gfrak$, these eigenvalues are determined by the roots of the algebra $\gfrak$. 
We will discuss this in the next section. 

%========================================
\subsection{Roots}

In this subsection we follow \cite{gilmore74}.
Since $X$ is a real antisymmetric matrix it can be diagonalized, and thus lies in the Cartan subalgebra. Let $r$ be the rank of the group $G$, which is equal
to the dimension of the Cartan subalgebra.
The generators 
$C_j$, $j=1,\dots, r$,
of the Cartan subalgebra in adjoint representation are $m\times m$ diagonal matrices of the form
\be
C_j=\diag\left(\underbrace{0,\dots,0}_{r},
i\alpha_{1j}, -i\alpha_{1j},\dots, 
i\alpha_{pj}, -i\alpha_{pj},
\right)\,,
\ee
where $\alpha_k=(\alpha_{kj})$, $k=1,\dots,p$, are some covectors in 
$\RR^r$ called the positive roots of the Lie algebra
$\gfrak$, $p$ is the number of positive roots related to the rank by
$r=m-2p$.
Then the matrix $X$ has the form
\be
X= \sum_{i=1}^{r}x^jC_j
=\diag\left(\underbrace{0,\dots,0}_{r},
i\alpha_1(x), -i\alpha_1(x),\dots, 
i\alpha_p(x), -i\alpha_p(x),
\right)\,,
\ee
where $x=(x^j)$, $j=1,\dots, r,$ is a vector in $\RR^r$, and 
$
\alpha_k(x)=\sum_{i=1}^{r}\alpha_{kj} x^j
$
is the canonical value of the covector $\alpha_k$ on the vector $x$ in $\RR^r$. 
The vector $x$ must be such 
that the value of each root on it is a half-integer, that is,
\be
\alpha_k(x)=\frac{n_k}{2}, \qquad n_k\in\ZZ\,, \qquad
k=1,\dots, p\,.
\ee
Thus, the number of (possibly) 
non-zero eigenvalues is equal to the number $p$
of positive roots of the algebra, and the eigenvalues themselves
are equal to the values of the roots on the vector $x$.

The roots $\alpha_k$ and the vector $x$ are vectors in $\RR^r$. However, it is convenient to consider this space $\RR^r$ as the hyperplane in $\RR^{r+1}$
orthogonal to the vector
$
b=\sum_{i=1}^{r+1}\hat e_i\,,
$
where $\hat e_i$, $i=1,\dots, r+1$, is the canonical orthonormal basis in $\RR^{r+1}$.
Then the roots $\alpha_k$ 
and the vector $x$ can be represented  in terms of the basis
$e_i$ of $\RR^{r+1}$ as
$
x=\sum_{i=1}^{r+1}\hat x^i \hat e_i\,,
$
and
$
\alpha_k = \sum_{i=1}^{r+1}\hat \alpha_{ki}\hat e_i\,,
$
$
k=1,\dots,p\,.
$
Notice that not all of the coordinates $\hat x^i$ are independent.
Since these vectors lie in the hyperplane orthogonal to the vector $b$,
the sum of all coordinates of these vectors should be equal to zero, that is,
$
\sum_{i=1}^{r+1}\hat x^i=0\,,
$
and 
$
\sum_{i=1}^{r+1}\hat \alpha_{ki}=0\,.
$
Then the values of the roots on the vector $x$ are
\be
\alpha_k(x)=\sum_{i=1}^{r+1}\hat \alpha_{ki}\hat x^i
=
\sum_{i=1}^{r}\left(\hat \alpha_{ki}
-\hat\alpha_{k,r+1}\right)\hat x^i
\,.
\ee
For all classical Lie algebras $A_n, B_n, C_n$ and $D_n$ the coordinates
of the roots are integers (up to a uniform normalization factor), that is,
$
\hat\alpha_{ki}=\lambda \beta_{ki}\,,
$
$k=1,\dots,p;\
i=1,\dots,r+1\,,
$
where $\beta_{ki}$ are integers and $\lambda$ is a normalization factor.
The normalization constant can be determined from the chosen Cartan-Killing metric (\ref{22xx}) by requiring
\be
\lambda^2\sum_{k=1}^p \beta_{ki}\beta_{kj}=\gamma_{ij}.
\ee
This can also be viewed as the definition of the metric
$\gamma_{ij}$. The constant $\lambda$ is then just a uniform factor
that can be set to $\lambda=1$.

Therefore, $\alpha_k(x)$ will be half-integer if 
$
\hat x^i=\frac{1}{2}k_i\,,
$
where $k_i$, $i=1,\dots, r,$ are arbitrary integers, that is,
\be
\alpha_k(x)=\frac{1}{2}
\sum_{i=1}^{r}\left(\beta_{ki}
-\beta_{k,r+1}\right)k_i\,.
\ee

To be specific let us consider the group $G=SU(N)$. 
The algebra $\mathfrak{su}(N)$ is isomorphic to the classical 
algebra $A_{N-1}$.
The dimension and the rank of $SU(N)$ are
\be
m=\dim SU(N)=N^2-1, \qquad r={\rm rank}\, SU(N)=N-1\,.
\ee
The positive roots
are labeled by two integers $1\le i<j\le N$, and have the form
$
\alpha_{ij}=\hat e_i-\hat e_j\,.
$
The number of positive roots is
\be
p=\frac{m-r}{2}=\frac{N(N-1)}{2}\,.
\ee
The Cartan-Killing metric is
$
\gamma_{ij}=2\delta_{ij}-\delta_{i,j-1}-\delta_{i,j+1}\,.
$
Then
\be
\alpha_{ij}(x)=\hat x^i-\hat x^j
=\frac{1}{2}(k_i-k_j)\,.
\ee

Therefore, in this case the integers determining the eigenvalues of the matrix $X$ are also labeled by two indices 
\be
n_{ij}=k_i-k_j\,, \qquad 1\le i<j\le N\,.
\ee
We will call these integers monopole numbers for $n_{ij}>0$ (or antimonopole numbers for $n_{ij}<0)$.
Here $k_i$, $i=1,\dots,N,$ are arbitrary integers whose sum is equal to zero. 
In particular, some or all of them can be equal to zero.
Note, however, that it is impossible to have only one non-zero number $k_i$. Therefore, either they are all equal to zero, or
there are at least two non-zero integers $k_i$. Another important observation is that for any choice of non-zero integers $k_i$ some of the integers $n_{ij}$ will have absolute value  greater or equal to $2$. In other words, it is impossible to have $n_{ij}=0,\pm 1$
for all $i,j$ (except, of course, the trivial case when all integers $k_i=0$). This observation has profound implications for the stability of the chromomagnetic vacuum studied in this paper.

For the classical groups the situation is similar.
Let, as above, $k_i$, $i=1,\dots,N,$ be an arbitrary collection of $N$ integers whose sum is equal to zero.
Then for the group $D_{N-1}$ 
the possible monopole numbers are
\be
n_{ij}=\pm k_i\pm k_j\,.
\ee
For the group $B_{N-1}$ there are two possible combinations
\be
n_{ij}=\pm k_i\pm k_j
\qquad\mbox{and}\qquad
n_i=\pm k_i\,,
\ee
and for the group $C_{N-1}$ the possible combinations are
\be
n_{ij}=\pm k_i\pm k_j
\qquad\mbox{and}\qquad
n_i=\pm 2k_i\,.
\ee
It is not difficult to see that in all these cases there is no choice of
non-zero integers $k_i$ such that the only monopole numbers are $0, \pm 1$. 
There will be necessarily monopole numbers with absolute value greater or equal to $2$.

Thus, {\it for any of the compact simple classical groups if the matrix $X$ has at least one non-zero eigenvalue, then it will have at least one eigenvalue with absolute value greater or equal to $2$}.

%==================================
\subsection{Weights}

Now let us consider an irreducible representation $\gfrak\to \End(W)$ of the Lie algebra $\gfrak$ in a $N$-dimensional
complex vector space $W$. The generators of the Cartan subalgebra in this representation,
$W(C_i)$, $i=1,\dots,r$, are $N\times N$ complex diagonal matrices of the
form
\be
W(C_j)=\diag\left(i\nu_{1j},\dots,i\nu_{Nj}\right)\,,
\ee
where $\nu_k=(\nu_{kj})$, $k=1,\dots,N$, are some covectors in 
$\RR^r$ called the weights of the representation $W$.
Contrary to roots, the weights can be degenerate, that is, have multiplicity greater than $1$, and be equal to zero with some multiplicity too.
Then the matrix $W(X)$ has the form
\be
W(X)=\sum_{i=1}^{r}x^jW(C_j)
=\diag\Big(i\nu_1(x), \dots, i\nu_N(x)
\Big)\,,
\ee
where $x=(x^j)$, $j=1,\dots, r,$ is a vector in $\RR^r$, and 
$
\nu_k(x)=\sum_{i=1}^{r}\nu_{kj} x^j
$
is the canonical value of the covector $\alpha_k$ on the vector $x$ in $\RR^r$. 
The vector $x$ must be such 
that the value of each weight on it is a half-integer, that is,
\be
\nu_k(x)=\frac{m_k}{2}\,,\qquad m_k\in\ZZ\,,
\qquad
k=1,\dots, N\,.
\ee

The weights lie in the same space as the roots. So, we can represent them
by
$
\nu_k = \sum_{i=1}^{r+1}\hat \nu_{ki}\hat e_i\,,
$\;
$
k=1,\dots,N\,,
$
where
$
\sum_{i=1}^{r+1}\hat \nu_{ki}=0\,.
$
Then the values of the weights on the vector $x$ are
\be
\nu_k(x)=\sum_{i=1}^{r+1}\hat \nu_{ki}\hat x^i
=\sum_{i=1}^{r}\left(\hat \nu_{ki}
-\hat\nu_{k,r+1}\right)\hat x^i
=\frac{1}{2}
\sum_{i=1}^{r}\left(\hat\nu_{ki}
-\hat\nu_{k,r+1}\right)k_i
\,,
\ee
where
$k_i$, $i=1,\dots, r,$ are arbitrary integers.

To be specific let us consider the {\it fundamental} (defining) representation of the algebra $\gfrak=\mathfrak{su}(N)$ by $N\times N$ complex traceless
anti-Hermitian matrices. The Cartan subalgebra is generated by diagonal matrices. Therefore, the generators $W(C_i)$, $i=1,\dots,N-1,$ of the Cartan subalgebra must have $N$ imaginary eigenvalues whose sum is equal to zero. Then the matrix $W(X)$ must have the form
\be
W(X)=\sum_{i=1}^{r}x^jW(C_j)
=\diag\left(i\frac{k_1}{2}, \dots, i\frac{k_N}{2}\right)\,,
\ee
where $k_i$, $i=1,\dots,N,$ are $N$ integers whose sum is equal to zero,
These are exactly the integers that define the values of the roots $\alpha_{ij}(x)=(k_i-k_j)/2$ for the adjoint representation of the group $SU(N)$.

%=======================
\subsection{Isometries}

It is well-known that $S^2=SO(3)/SO(2)$, so that $SO(3)$ is the isometry group
and $SO(2)$ is the isotropy (or holonomy) group of $S^2$.
The Killing vectors, $\xi_A$, of $S^2$ have the form
\cite{avramidi09}
\bea
\xi_1&=&\cos\varphi\,\partial_r
-\frac{1}{a}\cot(r/a)\sin\varphi\,\partial_\varphi\,,
\\
\xi_2&=&\sin\varphi\,\partial_r
+\frac{1}{a}\cot(r/a)\cos\varphi\,\partial_\varphi\,,
\\
\xi_3&=&\partial_\varphi\,.
\eea
One can check that the Killing vector fields form a representation of the
isometry algebra, $SO(3)$,
\cite{avramidi09}
\bea
[\xi_1, \xi_2]=-\frac{1}{a^2}\xi_{3}\,,
\qquad
{}[\xi_{3}, \xi_1]=-\xi_2
\qquad
{}[\xi_{2}, \xi_3]=-\xi_1\,.
\eea
The Cartan metric of the group $SO(3)$
has the form
\be
(\gamma_{AB})=\diag(1,1,a^2)\,,
\qquad
(\gamma^{AB})=\diag\left(1,1,\frac{1}{a^2}\right)\,.
\ee
Therefore, the Casimir operator, 
$\Delta = \gamma^{AB}\xi_A\xi_B$, of the Lie algebra 
of the group $SO(3)$ is
nothing but the scalar Laplacian
\bea
\Delta &=&
\partial_r^2+\frac{1}{a}\cot(r/a)\partial_r
+\frac{1}{a^2\sin^2(r/a)}\partial_\varphi^2\,.
\eea

Now, suppose the group $G$ has the spin group ${\rm Spin}(2)$
as a subgroup and let $\alpha: {\rm Spin}(2)\to G$ be the corresponding
embedding. Since we also have a representation $\rho_W: G\to {\rm Aut}(W)$ of the gauge group $G$ in the vector space $W$, this defines a new 
representation of the spin group 
$\rho_X=\rho_W\circ\alpha: {\rm Spin}(2)\to {\rm Aut}(W)$.
Let $\Sigma$ be the generator of the spin group ${\rm Spin}(2)$.
Then $X(\Sigma)$ is the generator of the spin group ${\rm Spin}(2)$
in the representation $\rho_X$
and
\be
G=\II_W\otimes T(\Sigma)-X(\Sigma)\otimes\II_T
\ee
is the generator of the twisted 
representation $X\otimes T$ of the spin group ${\rm Spin}(2)$.
This generator should not be confused with the gauge group denoted by the same symbol.

The twisted Lie derivatives ${\cal L}_A={\cal L}_{\xi_A}$ 
along Killing vectors $\xi_A$
of sections of the vector bundle ${\cal V}$ are
\cite{avramidi09}
\bea
{\cal L}_1&=&\cos\varphi\,\partial_r
-\frac{1}{a}\sin\varphi\cot(r/a)\,\partial_\varphi
%\nonumber\\
%&&
+\sin\varphi\frac{1-\cos(r/a)}{a\sin(r/a)} G
\,,
\\
%[10pt]
{\cal L}_2&=&\sin\varphi\,\partial_r
+\frac{1}{a}\cos\varphi\cot(r/a)\,\partial_\varphi
%\nonumber\\
%&&
-\cos\varphi\frac{1-\cos(r/a)}{a\sin(r/a)} G
\,,
\\
{\cal L}_3&=&
\partial_\varphi+G\,.
\eea
One can show that these operators form
a representation of the isometry algebra
$\mathfrak{so}(3)$
\cite{avramidi09}
\bea
[{\cal L}_1, {\cal L}_2]=-\frac{1}{a^2}{\cal L}_3\,,
\qquad
{}[{\cal L}_3, {\cal L}_1]=-{\cal L}_2
\qquad
{}[{\cal L}_2, {\cal L}_3]=-{\cal L}_1
\,.
\eea
The generalized Laplacian is expressed in terms of the 
Casimir operators of the isometry group $SO(3)$
and the holonomy group $SO(2)$
\be
\Delta=\gamma^{AB}{\cal L}_A{\cal L}_B-\frac{1}{a^2}G^2
\ee
and is equal to
\cite{avramidi09}
\bea
\Delta &=&
\partial_r^2+\frac{1}{a}\cot(r/a)\partial_r
+\frac{1}{a^2\sin^2(r/a)}
\left(\partial_\varphi
+[1-\cos(r/a)]G\right)^2\,.
\label{459}
\eea

We will be interested in the limit as $a\to \infty$ (when the curvature of the sphere vanishes, so, formally, $S^2\to \RR^2$) and $X\to \infty$, so that the magnetic field 
$
H=-X(\Sigma)/a^2
$
remains constant.
In this limit  the generator $G$ becomes $G\to a^2H\otimes \II_T$.
It is worth noting that this limit is only defined locally
in the South coordinate patch.
Therefore, the global bundle structure is destroyed in this limit.
Thus, in the following we will be working in the South coordinate patch
(for small $r$) in $S^2$, which approaches $\RR^2$ as $a\to \infty$.

Notice that as $a\to \infty$ the Killing vectors of the sphere $S^2$
become the Killing vectors of the Euclidean space $\RR^2$
and the Laplacian becomes
exactly the Laplacian with a constant magnetic field
$H$
in the Euclidean plane $\RR^2$.

%=====================================================
\section{Heat Traces }
\setcounter{equation}0

We will employ the algebraic methods for calculation of the heat kernel
developed in 
\cite{avramidi08,avramidi09,avramidi10a}.
To compute the heat trace
directly
we need to compute the spectrum of the Laplacian, its eigenvalues and their multiplicities. However, we can also compute the heat trace
as an integral of the heat kernel diagonal. 
Since on the sphere the heat kernel
diagonal is constant the heat kernel is simply proportional to the fiber 
trace of the heat kernel diagonal,
\be
\Tr \exp(-tL)=\vol(M) \tr \exp(-tQ)U^{\rm diag}(t)\,.
\ee
That is why, we need to compute the heat kernel diagonal of the Laplacian.

Next, we note that on the product manifold $S^1\times S^1\times S^2$ the Laplacian splits naturally
\be
\Delta_{S^1\times S^1\times S^2}=
\Delta_{S^1\times S^1}+\Delta_{S^2}\,,
\ee
and, therefore, the heat kernel factorizes
\be
U^{\rm diag}_{S^1\times S^1\times S^2}(t)=
U^{\rm diag}_{S^1\times S^1}(t)
U^{\rm diag}_{S^2}(t)\,.
\ee
Therefore,
\be
\Tr \exp(-tL)=\vol(M) \tr \exp(-tQ)
U^{\rm diag}_{S^1\times S^1}(t)
U^{\rm diag}_{S^2}(t)\,.
\ee
That is why, we need to compute the heat kernel diagonals of the Laplacian
on $S^1$ and $S^2$ only.

%===========================================================
\subsection{Heat Kernel on $\RR^2$ and $S^1\times S^1$ without Magnetic Field}

The Laplacian on $\RR$ is
$
\Delta=\partial_x^2\,.
$
The heat kernel diagonal of such an operator is easily computed by Fourier
transform
\be
U^{\rm diag}_{\RR}(t)=\frac{1}{\sqrt{4\pi t}}\,.
\ee
For $\RR^2$ we obviously have the product
\be
U^{\rm diag}_{\RR^2}(t)=\frac{1}{4\pi t}\,.
\ee

On the circle $S^1$ of radius $a$ the heat kernel can be computed by Fourier
expansion.
The spectrum of the operator $(-\Delta)$ is
\be
\lambda_l(\Delta)=\frac{l^2}{a^2}\,,
\ee
where $l=0,1,2,\dots$, 
with multiplicities $d_0=1$ and $d_l=2$ for $l=1,2,\dots$. 
Then the heat kernel diagonal on $S^1$ is
\be
U^{\rm diag}_{S^1}(t)=
\frac{1}{\sqrt{4\pi t}}\Omega\left(\frac{t}{a^2}\right)
\,,
\ee
where
\bea
\Omega(t)
&=&
\sqrt{\frac{t}{\pi}}\left\{1+2\sum_{l=1}^\infty
e^{-tl^2}\right\}
=\sqrt{\frac{t}{\pi}}\,\theta_3\left(0,e^{-t}\right)
\,,
\eea
and $\theta_3(v,q)$ is the third Jacobi theta function.
Therefore, for the product $S^1\times S^1$ with radii $a_1$, $a_2$ the heat kernel diagonal is
\bea
U^{\rm diag}_{S^1\times S^1}(t)
&=&\frac{1}{4\pi t}
\Omega\left(\frac{t}{a_1^2}\right)
\Omega\left(\frac{t}{a_2^2}\right)
\,.
\eea

An important property of $\Omega(t)$ is the Poisson duality formula, which 
gives a nontrivial relation
\bea
\Omega(t)=
\sqrt{\frac{t}{\pi}}\,\Omega\left(\frac{\pi^2}{t}\right)
=\theta_3\left(0,e^{-\pi^2/t}\right).
\label{614xx}
\eea
By differentiating the equation (\ref{614xx}) we get
\be
\Omega'(t)=\frac{2\pi^2}{t^2}
\sum_{l=1}^\infty l^2 \exp\left(-\frac{\pi^2}{t}l^2\right)\,,
\ee
which immediately shows that $\Omega$ is an increasing function.

We will need the asymptotics of this function.
It is not difficult to see that
as $t\to 0$
\bea
\Omega(t)&=&1+2\exp\left(-\frac{\pi^2}{t}\right)
+O\left(e^{-4\pi^2/t}\right)\,,
\\
\Omega'(t)&=&2\pi^2 t^{-2}\exp\left(-\frac{\pi^2}{t}\right)
+O\left(e^{-4\pi^2/t}\right)\,,
\\
\Omega''(t)&=&2\pi^2 t^{-4}
\left(\pi^2-2t\right)
\exp\left(-\frac{\pi^2}{t}\right)
+O\left(e^{-4\pi^2/t}\right)\,,
\eea
and as $t\to \infty$, we have
\bea
\Omega(t)&=&\frac{1}{\sqrt{\pi}}\left[
t^{1/2}+2t^{1/2}e^{-t}
+O\left(e^{-4t}\right)\right]\,,
\\
\Omega'(t)&=&
\frac{1}{\sqrt{\pi}}
\left[
\frac{1}{2}t^{-1/2}
+\left(t^{-1/2}-2t^{1/2}\right)e^{-t}
+O\left(e^{-4t}\right)
\right]\,,
\\
\Omega''(t)&=&\frac{1}{\sqrt{\pi}}\left[
-\frac{1}{4}t^{-3/2}
+\left(2t^{1/2}-2t^{-1/2}-\frac{1}{2}t^{-3/2}\right)e^{-t}
+O\left(e^{-4t}\right)\right]\,.
\eea

Notice that although, in general, as $t\to \infty$ 
the derivatives of the function $\Omega(t)$ decrease only as powers of $t$,
a particular combination of the derivatives, which we will need later,
is exponentially decreasing as $t\to\infty$,
\bea
\Omega'(t)+2t\Omega''(t)&=&
\frac{2}{\sqrt{\pi}}t^{1/2}\left(2t-3\right)e^{-t}
+O\left(e^{-4t}\right)\,.
\label{888xx}
\eea
This will have
a major impact on the calculation of the heat capacity of the quark-gluon gas.

%======================================
\subsection{Heat Kernel on $\RR^2$ with Magnetic Field}

Let us now compute the heat kernel diagonal of the Laplacian with a constant
magnetic field on $\RR^2$. In the flat space limit
the Laplacian on $\RR^2$ in polar coordinates
with a constant magnetic field 
can be written as
\be
\Delta=\nabla_1^2+\nabla_2^2\,,
\ee
where
\bea
\nabla_{1}&=&\cos\varphi\,\partial_r
-\frac{1}{r}\sin\varphi\,\partial_\varphi
-\frac{r}{2}\sin\varphi H\,,
\\
\nabla_{2}&=&\sin\varphi\,\partial_r
+\frac{1}{r}\cos\varphi\,\partial_\varphi
+\frac{r}{2}\cos\varphi H\,.
\eea

The heat kernel can be computed as follows. First,
by observing that the covariant derivatives 
form the Heisenberg algebra
\be
[\nabla_1, \nabla_2]=H, \qquad
[\nabla_1,H]=[\nabla_2,H]=0
\,,
\ee
one can prove that the heat
semigroup can be expressed in the form
\cite{avramidi93}
\bea
\exp(t\Delta)&=&
\frac{1}{4\pi}
\frac{H}{\sin(tH)}
%\nonumber\\
%&&\times
\int\limits_{\RR^2}dq\;
\exp\left(-\frac{1}{4}H\cot(tH)|q|^2
+\left<q,\nabla\right>\right),
\nonumber\\
\eea
where $|q|^2=(q^1)^2+(q^2)^2$ and 
$\left<q,\nabla\right>=q^1\nabla_1+q^2\nabla_2$.
Further, it is not difficult to show that
\be
\left[\exp\left<q,\nabla\right>\delta(x,x')\right]^{\rm diag}
=\delta(q)\,,
\ee
so that the integral over $q$ becomes trivial and
we immediately obtain the heat kernel diagonal \cite{avramidi93}
\be
U^{\rm diag}_{\RR^2}(t)=\frac{1}{4\pi}\frac{H}{\sin(tH)}
\,.
\ee

Of course, since $\RR^2$ is non-compact, the spectrum of the Laplacian 
$\Delta$ is degenerate, that is, even if the eigenvalues are discrete
their multiplicities are infinite.
Therefore, the heat trace in the limit
$a\to \infty$ is infinite. However, the heat kernel diagonal is still
well defined.
Therefore, on the sphere $S^2$ in the limit as $a\to \infty$ and 
fixed $H=-X/a^2$ the heat kernel diagonal locally
must have the following limit
\be
\lim_{a\to \infty}U^{\rm diag}_{S^2}(t)
=\frac{1}{4\pi}\frac{H}{ \sin(tH)}
\,.
\ee
Recall that $H$ is a real anti-symmetric matrix with purely imaginary eigenvalues, 
so that this heat kernel diagonal is well defined.

%=============================================================
\subsection{Heat Trace on $S^2$}

\subsubsection{Algebraic Approach}

Let $(q^1,q^2,\omega)$ be the canonical coordinates on the isometry group
$SO(3)$.
Let $C$ be the contour of integration in the complex plane of $\omega$
defined by 
\be
C=\frac{1}{2}(C_++C_-)\,.
\ee
By using the isometry algebra and the representation of the Laplacian
in terms of the twisted Lie derivatives one can show that
the heat semigroup $\exp(t\Delta)$
can be represented in form of an integral over the isometry group 
\cite{avramidi09}
\bea
\exp(t\Delta) 
&=&
\frac{1}{4\pi t}
\exp\left\{
\left(\frac{1}{4}-G^2\right)\frac{t}{a^2}
\right\}
\int\limits_{C} 
\frac{d\omega}{\sqrt{4\pi t/a^2}}\;
\int\limits_{\RR^2}dq\;
\exp\left\{
-\frac{|q|^2}{ 4t}
-\frac{a^2}{4t}\omega^2
\right\}
\nonumber\\
%&&
%\times
%\nonumber
&&
\times
\frac{\sin\left[\omega/2\right]}{\omega/2}
\exp\left[\left<q,\mathcal{L}\right>
+\omega{\cal L}_3\right]
\,.
\label{49}
%\nonumber
\eea
where $|q|^2=(q^1)^2+(q^2)^2$, $\left<q,{\cal L}\right>=q^1\mathcal{L}_1
+q^2\mathcal{L}_2$.
We will see later that one can take here either $C_+$ or $C_-$, which gives identical
results.

By using the isometry algebra one can show that
\cite{avramidi09}
\be
\exp\left[\left<q,\mathcal{L}\right>+\omega{\cal L}_3\right]
\delta(x,x')\Big|_{x=x'}
=\left(\frac{\omega/2}
{\sin\left[\omega/2\right]}\right)^{2}
\exp\left(\omega G\right)
\delta(q)\,.
\ee
Substituting this into (\ref{49}) 
we obtain the heat kernel diagonal in the form
\bea
U^{\rm diag}(t) 
&=&\frac{1}{4\pi t}
\exp\left[\frac{t}{a^2}\left(\frac{1}{4}-G^2\right)\right]
\int\limits_{C}
\frac{ad\omega}{\sqrt{4\pi t}}\;
\exp\left\{-\frac{a^2\omega^2}{4t}+G\omega\right\}
%\exp(G\omega)
\frac{\omega/2}
{\sin\left[\omega/2\right]}\,.
\nonumber\\
\label{5139}
\eea
This can be written in the form
\bea
U^{\rm diag}(t) 
&=&\frac{1}{4\pi t}
\exp\left[\frac{t}{a^2}\left(\frac{1}{4}-G^2\right)\right]
\Psi\left(\frac{t}{a^2};-2iG\right)\,,
%\label{5139}
\eea
where $\Psi(t;n)$, with an integer $n$,  is a function defined by
\bea
\Psi(t;n)&=&
\int\limits_{C}
\frac{d\omega}{\sqrt{4\pi t}}\;
\exp\left\{-\frac{\omega^2}{4t}+in\omega/2\right\}
\frac{\omega/2}
{\sin\left[\omega/2\right]},
\label{5214}
\eea
Recall that $G$ has imaginary half-integer eigenvalues.

%=============================================================
\subsubsection{Properties of the Function $\Psi(t;n)$}

Let
\bea
\Psi_\pm(t;n)&=&
\int\limits_{C_\pm}
\frac{d\omega}{\sqrt{4\pi t}}\;
\exp\left\{-\frac{\omega^2}{4t}+in\omega/2\right\}
\frac{\omega/2}
{\sin\left[\omega/2\right]}\,,
%\label{5214}
\eea
so that
\bea
\Psi(t;n)
&=&\frac{1}{2}\left[\Psi_+(t;n)+\Psi_-(t;n)\right]
\,.
\eea
By deforming the contour of integration one can show that
\bea
\Psi_-(t;n)&=&\Psi_+(t;n)+i[R_-(t;n)+R_+(t;n)]\,,
\eea
where
\bea
R_+(t;n)&=&
\left(\frac{\pi}{t}\right)^{1/2}
\sum_{k=1}^\infty (-1)^{k(1+n)}2\pi k
\exp\left\{-\frac{\pi^2k^2}{t}\right\}\,,
\\
R_-(t;n)&=&
\left(\frac{\pi}{t}\right)^{1/2}
\sum_{k=-1}^{-\infty} (-1)^{k(1+n)}2\pi k
\exp\left\{-\frac{\pi^2k^2}{t}\right\}\,.
\eea
Obviously,
\be
R_-(t;n)=-R_+(t;n)\,.
\ee
Therefore,
\be
\Psi(t;n)=\Psi_+(t;n)=\Psi_-(t;n)\,,
\ee
and

%\end{document}

\bea
\Psi(t;n)
&=&
\fint\limits_{-\infty}^\infty
\frac{d\omega}{\sqrt{4\pi t}}\;
\exp\left\{-\frac{\omega^2}{4t}\right\}
\frac{\omega/2}
{\sin\left[\omega/2\right]}\cos(n\omega/2)
\,,
\label{544}
\eea
where $\fint$ denotes the Cauchy principal value of the integral. 
This can also be written as
\bea
\Psi(t;n)
&=&
\sum_{k=-\infty}^\infty
(-1)^{k(1+n)}
\fint\limits_{0}^{2\pi}
\frac{d\omega}{\sqrt{4\pi t}}\;
\exp\left\{-\frac{1}{4t}(\omega+2\pi k)^2\right\}
\nonumber\\
&&\times
\frac{(\omega+2\pi k)/2}
{\sin\left[\omega/2\right]}\cos(n\omega/2)
\,,
%\nonumber\\
%\label{5139}
\eea
which is nothing but the sum over closed geodesics of $S^2$.

We can also rewrite the function $\Psi(t;n)$ in the form
\bea
\Psi(t;n)&=&
\fint\limits_{-\infty}^\infty
\frac{d\omega}{\sqrt{4\pi}}\;
\exp\left\{-\frac{\omega^2}{4}\right\}
\frac{\omega\sqrt{t}/2}
{\sin\left[\omega\sqrt{t}/2\right]}\cos\left[n\omega\sqrt{t}/2\right]\,,
\label{530xxz}
\eea
which allows the analytic continuation in the complex plane of $t$ with a cut
along the negative real axis (so that $t=|t|e^{i\theta}$ with $|\theta|<\pi$),
\bea
\Psi(t;n)&=&
\int\limits_{-\infty}^\infty
\frac{d\omega}{\sqrt{4\pi}}\;
\exp\left\{-\frac{\omega^2}{4}\right\}
\frac{\omega\sqrt{-t}/2}
{\sinh\left[\omega\sqrt{-t}/2\right]}
\cosh\left[n\omega\sqrt{-t}/2\right]\,.
%\label{5214}
\eea

Note also that the function $\Psi(t;n)$ is an even function of $n$, that is,
\be
\Psi(t;n)=\Psi(t;-n)=\Psi(t;|n|)\,,
\ee
so that it depends only on the absolute value
$|n|$. In terms of the matrix $G$ this means that
$\Psi(t;-2iG)$ depends only on the absolute value $|G|$ of the matrix $G$, 
which can be defined as the positive square root of the real
symmetric matrix $|G|=\sqrt{-G^2}$. Recall that $G$ has purely imaginary half-integer eigenvalues, therefore, $|G|$ has positive half-integer eigenvalues. 

We can also find the dual representation of the function $\Psi(t;n)$
as follows.
As we have seen above one can use either contour, $C_+$ or $C_-$ for the calculation
of the function $\Psi(t;n)$. We use the series
the series 
\be
\frac{\omega/2}{\sin\left[\omega/2\right]}
=-i\omega\sum_{l=0}^\infty
\exp\left\{
i\left(l+\frac{1}{2}\right)\omega
\right\}
\ee
for ${\rm Im}\,\omega>0$ and
\be
\frac{\omega/2}{\sin\left[\omega/2\right]}
=
i\omega\sum_{l=0}^\infty
\exp\left\{
-i\left(l+\frac{1}{2}\right)\omega
\right\}
\ee
for ${\rm Im}\,\omega<0$; they both 
converge absolutely. 

Now we notice that
\bea
\int\limits_{C_+-C_-} \frac{d\omega}{\sqrt{4\pi t}}
\exp\left\{-\frac{\omega^2}{4t}\right\}
\frac{\omega/2}{\sin\left[\omega/2\right]}\sin[n\omega/2]=0\,.
\eea
Indeed, the integrand is analytic on the real axis
and, therefore, this integral vanishes.
Therefore, we can add it to the original integral to get
\bea
\Psi(t;n)
&=&
\frac{1}{2}\int\limits_{C_++C_-} \frac{d\omega}{\sqrt{4\pi t}}
\exp\left\{-\frac{\omega^2}{4t}\right\}
\frac{\omega/2}{\sin\left[\omega/2\right]}\cos[n\omega/2]
\nonumber\\
&&
+
\frac{i}{2}\int\limits_{C_+-C_-} \frac{d\omega}{\sqrt{4\pi t}}
\exp\left\{-\frac{\omega^2}{4t}\right\}
\frac{\omega/2}{\sin\left[\omega/2\right]}\sin[|n|\omega/2]
\nonumber\\
&=&
\frac{1}{2}\int\limits_{C_+} \frac{d\omega}{\sqrt{4\pi t}}
\exp\left\{-\frac{\omega^2}{4t}+i|n|\omega/2\right\}
\frac{\omega/2}{\sin\left[\omega/2\right]}
\nonumber\\
&&
+
\frac{1}{2}\int\limits_{C_-} \frac{d\omega}{\sqrt{4\pi t}}
\exp\left\{-\frac{\omega^2}{4t}-i|n|\omega/2\right\}
\frac{\omega/2}{\sin\left[\omega/2\right]}\,.
\nonumber\\
\eea
Now, by using the expansion above we get
\bea
\Psi(t;n)
&=&
\frac{1}{2}\int\limits_{C_+} \frac{d\omega}{\sqrt{4\pi t}}
\exp\left\{-\frac{\omega^2}{4t}+i|n|\omega/2\right\}
\sum_{l=0}^\infty
(-i\omega)\exp\left\{i\left(l+\frac{1}{2}\right)\omega\right\}
\nonumber\\
&&
+
\frac{1}{2}\int\limits_{C_-} \frac{d\omega}{\sqrt{4\pi t}}
\exp\left\{-\frac{\omega^2}{4t}-i|n|\omega/2\right\}
\sum_{l=0}^\infty
i\omega\exp\left\{-i\left(l+\frac{1}{2}\right)\omega\right\}\,.
\nonumber\\
\eea
Finally, by using the equation
\bea
&&\mp\int\limits_{C_\pm}
\frac{d\omega}{\sqrt{4\pi t}}
\exp\left\{
-\frac{\omega^2}{4t}\pm i\left(l+\frac{1+|n|}{2}\right)\omega
\right\}i\omega
\nonumber\\
&&
\qquad\qquad
=2t\left(l+\frac{1+|n|}{2}\right)
\exp\left\{-t\left(l+\frac{1+|n|}{2}\right)^2\right\}\,,
%\nonumber\\
\eea
we obtain the spectral representation
\be
\Psi(t;n)=t\sum_{l=0}^\infty 
2\left(l+\frac{1+|n|}{2}\right)
\exp\left\{-t\left(l+\frac{1+|n|}{2}\right)^2\right\}\,.
\ee

By using this equation we see that all functions $\Psi(t;n)$ for different $n$
can be reduced to just two functions, 
$\Psi(t;0)$ and $\Psi(t;1)$, namely, for $n\ge 0$
\bea
\Psi(t;2n)&=&
\Psi(t;0)
-t\sum_{l=0}^{n-1}
\left(2l+1\right)
\exp\left\{-t\left(l+\frac{1}{2}\right)^2\right\}\,,
\\
\Psi(t;2n+1)&=&
\Psi(t;1)
-2t\sum_{l=1}^{n}
le^{-tl^2}\,.
\eea
Also, it is not difficult to see that the function $\Psi(t;0)$ can be expressed in terms of the function $\Psi(t;1)$ as follows
\be
\Psi(t;0)=2\Psi\left(\frac{t}{4};1\right)-\Psi(t;1)\,.
\ee
This property can be obtained directly from the integral representation (\ref{544}).

One can also show that the function $\Psi(t;n)$ has the following dual 
integral representation
\bea
\Psi(t;n)&=&t\int\limits_{\Gamma}
d\nu\;\tan(\pi\nu)
i\left(\nu+\frac{|n|}{2}\right)
\exp\left[-t\left(\nu+\frac{|n|}{2}\right)^2\right]\,,
\eea
where
$\Gamma$ is a contour that goes counterclockwise from $(\infty+i\varepsilon)$ 
to $(\frac{1}{2}+i\varepsilon)$, then around the point $\frac{1}{2}$
between $0$ and $\frac{1}{2}$ on the real axis 
to the point $(\frac{1}{2}-i\varepsilon)$, and finally
to $(\infty-i\varepsilon)$.

%===============================================================
\subsubsection{Heat Trace and the Spectrum}

By using the spectral representation of the function $\Psi(t;n)$
we can write the heat kernel diagonal of the Laplacian in the form
\bea
U^{\rm diag}(t) 
&=&\frac{1}{4\pi a^2}
\sum_{l=0}^\infty 
2\left(l+\frac{1}{2}+|G|\right)
\exp\left\{-\frac{t}{a^2}
\left[\left(l+\frac{1}{2}+|G|\right)^2
-\frac{1}{4}+G^2
\right]\right\}
\nonumber\\ 
\label{655}
&=&\frac{1}{4\pi a^2}
\sum_{l=0}^\infty d_l\left(2|G|\right) 
\exp\left\{-\frac{t}{a^2}\lambda_l\left(2|G|\right)\right\}
\,,
\eea
where
\bea
\lambda_l(n)&=&
l(l+n+1)+\frac{n}{2}\,,
\label{541xx}
\\
d_l(n)&=&2l+n+1\,.
\label{542xx}
\eea
Then the heat trace of the Laplace type operator
$L=-\Delta+Q$ is 
\bea
\Tr\exp(-tL)
&=&
\tr
\sum_{l=0}^\infty
d_l\left(2|G|\right) 
\exp\left\{-\frac{t}{a^2}\left[\lambda_l\left(2|G|\right)+a^2Q\right]\right\}
\,.
\label{658}
\eea

Recall that $G$ is a generator of $SO(2)$ in some $N$-dimensional
representation. We can diagonalize 
as follows
\be
G=\diag\left(i\frac{m_1}{2},\dots, i\frac{m_N}{2}\right)\,,
\ee
where $m_1, \dots, m_N$ are some integers.
Since the matrix $Q$ commutes with the matrix $G$, it can be diagonalized
simultaneously with $G$, that is,
\be
Q=\diag\left(Q_1,\dots,Q_N\right)\,.
\ee

By using this decomposition the heat trace takes the form
\bea
\Tr\exp(-tL)
&=&
\sum_{j=1}^N
\sum_{l=0}^\infty
d_{l}(|m_j|)
\exp\left(-\frac{t}{a^2}\lambda_{l,j}\right)
\,,
\eea
where
\bea
\lambda_{l,j}&=&
\lambda_l(|m_j|)+a^2 Q_j\,,
\qquad j=1,\dots,N,
\,.
\eea
Therefore, the eigenvalues of the operator $L$ are
$\frac{1}{a^2}\lambda_{l,j}$\,, $j=1,\dots, N;$ $l=0,1,2,\dots, $
with multiplicities $d_l(|m_j|)$.
The smallest eigenvalue of the operator $L$ is
\be
\lambda_{\rm min}(L)=\frac{1}{a^2}\min_{1\le j\le N}
\left\{\lambda_0(|m_j|)+Q_j\right\}
=\min_{1\le j\le N}
\left\{\frac{1}{a^2}\frac{|m_j|}{2}
+Q_j\right\}\,.
\ee

In terms of the function $\Psi(t;n)$ the heat trace
can be written as
\bea
\Tr\exp(-tL)
&=&
\frac{a^2}{t}
\sum_{j=1}^N
\exp\left[\frac{t}{a^2}\left(\frac{(1+m_j^2)}{4}-a^2Q_j\right)\right]
\Psi\left(\frac{t}{a^2};m_j\right)
\,.
\nonumber\\
\eea

%======================
\subsubsection{Asymptotics}
%======================

The asymptotics of the function $\Psi(t;n)$ as $t\to \infty$ can be directly obtained 
from the eigenvalue representation; it is given by the lowest eigenvalue
\be
\Psi(t;n)\sim \left(1+|n|\right)t
\exp\left\{-t\frac{(1+|n|)^2}{4}\right\}
+\left(3+|n|\right)t
\exp\left\{-t\frac{(3+|n|)^2}{4}\right\}
+\cdots
\,.
\ee

To compute the asymptotics as $t\to 0$ we use the rescaled representation
(\ref{530xxz}).
This form is particularly useful to compute the short-time asymptotics
as $t\to 0$; we just expand the exponent in the powers of $t$
and compute the Gaussian integrals over $\omega$.
The asymptotics as $t\to 0$ are
\be
\Psi(t;n)\sim\sum_{k=0}^\infty c_k(n)t^k,
\ee
where
\bea
c_0(n)&=&1,\qquad
\\
c_1(n)&=&\frac{1}{12} - \frac{n^2}{4},\qquad
\\
c_2(n)&=&\frac{7}{480} - \frac{n^2}{16} + \frac{n^4}{32}\,,
\\
c_3(n) &=& \frac{31}{8064}
-\frac{7n^2}{384}
+\frac{5n^4}{384}
-\frac{n^6}{384}\,.
\eea
Therefore, the heat kernel diagonal asymptotics as $t\to 0$ are
\be
U^{\rm diag}(t)\sim (4\pi t)^{-1}\sum_{k=0}^\infty a_k\left(\frac{t}{a}\right)^k,
\ee
where
\be
a_0=1, \qquad
a_1=\frac{1}{3},\qquad
a_2=\frac{1}{15}+\frac{1}{6}G^2\,.
\ee
This coincides with the well-known 
general heat kernel coefficients (\ref{317}).

Moreover, this integral can be used to compute the 
asymptotics of the heat kernel diagonal
as $a\to\infty$. It amounts to just
shifting the contour of integration in (\ref{5139})
by $\omega\mapsto\omega+2Gt/a^2$ to get
\bea
U^{\rm diag}(t)
&=&\frac{1}{4\pi t}
\exp\left(\frac{t}{4a^2}\right)
\int\limits_{-\infty}^\infty
\frac{d\omega}{\sqrt{4\pi}}\;
\exp\left(-\frac{\omega^2}{4}\right)
%\nonumber\\
%&&\times
\frac{\omega\sqrt{t}/(2a)+Gt/a^2}
{\sin\left[\omega\sqrt{t}/(2a)+Gt/a^2\right]}
\nonumber\\
%\label{5141}
\eea
Next, let $G=T+a^2H$, where $T=T(\Sigma)$ is a generator of the
group $SO(2)$ in the representation $T$ and $H=-X(\Sigma)/a^2$ is the 
generator of the group $SO(2)$ in the representation $X$ 
taking values in the adjoint representation of the Lie algebra of the
gauge group. Then
\bea
U^{\rm diag}(t) 
&=&\frac{1}{4\pi t}
\exp\left(\frac{t}{4a^2}\right)
\int\limits_{-\infty}^\infty\frac{d\omega}{\sqrt{4\pi}}\;
\exp\left(-\frac{\omega^2}{4}\right)
%\nonumber\\
%&&\times
\frac{tH+\omega\sqrt{t}/(2a)+tT/a^2}
{\sin\left[tH+\omega\sqrt{t}/(2a)+tT/a^2\right]}
\,.
\nonumber\\
\eea

Now, the asymptotics as $a\to \infty$ can be computed
by expanding the integrand in a Taylor series in
inverse powers of $a$ and computing the Gaussian integrals over $\omega$.
This gives the correct leading asymptotics that coincides with the
Euclidean space limit (with the magnetic field $H$)
\bea
\lim_{a\to \infty}U^{\rm diag}(t) 
&=&\frac{1}{4\pi}\frac{H}
{\sin\left(tH\right)}
\,.
\label{5140}
\eea
Of course, such asymptotics describe the behavior of the heat kernel for large $a$ and fixed $t$. The behavior of the heat kernel for large $t$ is sensitive to 
the value of $a$, it is described by the lowest eigenvalue of the Laplacian, which
is different on the sphere $S^2$ and on the Euclidean space $\RR^2$.

%=============================================
\subsubsection{Spectral Approach}
%=============================================

As an alternative, let us compute the spectrum of the Laplacian directly.
The Laplacian (\ref{459}) is a differential operator
acting on the space
$L^2([0,a\pi]\times[0,2\pi],a\sin(r/a)dr\,d\varphi)$.
We will find it useful to introduce a new variable 
$
x = \cos(r/a)\,.
$
Then the Laplacian acts on the space
$L^2([-1,1]\times[0,2\pi],a^2dx\,d\varphi)$
and takes the form
\be
\Delta
=\frac{1}{a^2}\left\{(1-x^2)\partial^2_x
-2x\partial_x
+\frac{1}{1-x^2}\left[\partial_\varphi+(1-x)G\right]^2\right\}\,.
\ee

Let $\alpha, \beta$ be {\it positive}
Hermitian operators defined by
\bea
\alpha=
\left|
\partial_\varphi
\right|
\,,\qquad
\beta=
\left|\partial_\varphi+2G\right|\,,
\eea
and
\be
\rho(x)=(1-x)^\alpha(1+x)^\beta\,.
\ee
We intertwine the Laplacian as follows
\bea
\tilde\Delta
&=&\rho^{-1/2}\Delta\rho^{1/2}
\nonumber\\
&=&-\frac{1}{a^2}\left(K_{\alpha,\beta}
+\frac{1}{4}\left(\alpha+\beta+1\right)^2
-\frac{1}{4}
+G^2
\right)\,,
\eea
where
\bea
K_{\alpha,\beta}&=&-(1-x^2)\partial^2_x
-\left[\beta-\alpha-(\alpha+\beta+2)x\right]\partial_x
\,.
\eea

Then the operator $\tilde\Delta$ has the same spectrum and, therefore,
the same heat trace as the original Laplacian.
The operator $K_{\alpha,\beta}$ is an
elliptic self-adjoint operator on $L^2([-1,1]\times[0,2\pi],\rho)$
with the weight function  $\rho(x)$.
It is well-known 
\cite{nikiforov88} that the operator $K_{\alpha,\beta}$
(with positive
real $\alpha,\beta>0$)
has a discrete spectrum with simple eigenvalues
\be
\lambda_l(K_{\alpha,\beta})=l(l+\alpha+\beta+1)\,,
\ee
where $l=0,1,\dots$, with eigenfunctions proportional to the
Jacobi polynomials
$P^{(\alpha,\beta)}_l(x)$.

This immediately enables us to compute the heat trace of a Laplace type operator $L=-\Delta+Q$ as follows
\bea
\Tr\exp(-tL)
&=&
\sum_{l=0}^\infty
\Tr
\exp\left\{-\frac{t}{a^2}\left(
\left(l+\frac{1}{2}(\alpha+\beta+1)\right)^2
+G^2
-\frac{1}{4}
+a^2Q\right)\right\}\,,
\nonumber\\
\eea
where the trace $\Tr$ in the last equation is the combined trace over
the operator $\partial_\varphi$ and the fiber trace over the matrix $G$.
Since the eigenvalues of the operator $\partial_\varphi$ are 
purely imaginary integers, $im$, with $m\in\ZZ$,
this can finally be written as 
\bea
\Tr\exp(-tL)
&=&
\tr
\exp\left\{-\frac{t}{a^2}\left(
G^2
-\frac{1}{4}
+a^2Q\right)\right\}
\nonumber\\
&&\times
\sum_{l=0}^\infty\sum_{m=-\infty}^\infty
\exp\left\{-\frac{t}{a^2}\left[
l+\frac{1}{2}\left(|m|+|m-2iG|+1\right)
\right]^2\right\}\,,
\eea
where the trace $\tr$ is now over the fiber indices only.

Now, one can show that this sum is equal to the previous result
(\ref{655}), (\ref{658}),
obtained by a completely different algebraic approach.

%======================================================
\section{Effective Action}
\setcounter{equation}0

%===============================================
\subsection{Yang-Mills Effective Action}

The magnitude of the magnetic field in our scenario can be written as 
$|{\cal F}|^2=\frac{1}{4}K/a^4$, 
where 
$
K=\sum_{i=1}^p n_i^2\,,
$
with $p$ being the number of positive roots of the gauge algebra, and,
therefore,
the classical (Euclidean) 
action on the covariantly constant background $S^1\times S^1\times S^2$ is
\be
S=16\pi^3\sigma
\frac{1}{xy}
\,,
\ee
where $\sigma=K/8e^2$,
\be
x=\frac{a}{a_1},\qquad
y=\frac{a}{a_2}\,,
\ee
and $a_1$ and $a_2$ are the radii of the two circles.
The total effective action is given by the sum of this classical part and the one-loop effective action (\ref{316}).

For the Yang-Mills ghost operator, the endomorphism $Q$ and the generator $G$ are
\be
Q^{YM}_{\rm ghost}=0, \qquad G^{YM}_{\rm ghost}=-X\,,
\ee
where $X$ is the generator of the spin group ${\rm Spin}(2)$ 
taking values in the adjoint representation of the Lie group $G$.
The volume of the manifold $M=S^1\times S^1\times S^2$
is $\vol(M)=16\pi^3 a^2a_1a_2$.
By taking into account the contribution of the torus $S^1\times S^1$, the heat trace of the Yang-Mills ghost operator is
\bea
\Theta^{YM}_{\rm ghost}(t) 
&=& 16\pi^3 a^2a_1a_2\Omega\left(\frac{t}{a_1^2}\right)
\Omega\left(\frac{t}{a_2^2}\right)
\tr_{Ad}
\exp\left[\frac{t}{a^2}\left(\frac{1}{4}-X^2\right)\right]
\Psi\left(\frac{t}{a^2};2iX\right)\,.
%\label{5139}
\eea

For the Yang-Mills operator $L_{\rm vect}$ the endomorphism $Q_{\rm vect}$ is
given by
\be
a^2Q_{\rm vect}=P+2EX\,,
\ee
where $P$ and $E$ are $4\times 4$ real matrices of the form
\bea
P^a{}_b=\delta^a{}_2\delta_{2b}
+\delta^a{}_3\delta{}_{3b}\,,
\qquad
%\\
E^a{}_b=\delta^a{}_2\delta_{3b}
-\delta^a{}_3\delta_{2b}
\,.
%\qquad
\eea
Notice that $P$ is a $2$-dimensional projection, that is, $P^2=P$, and $E^2=-P$.
The generator $G_{\rm vect}$ is
\be
G_{\rm vect}=E-X\,.
\ee
Therefore, 
\be
G_{\rm vect}^2+a^2Q_{\rm vect}=X^2.
\ee
So, again, by adding the corresponding factors for the torus
$S^1\times S^1$ we get the heat kernel diagonal of the Yang-Mills
operator
\bea
\Theta_{\rm vect}(t) 
&=&16\pi^3 a^2a_1a_2
\Omega\left(\frac{t}{a_1^2}\right)
\Omega\left(\frac{t}{a_2^2}\right)
\tr_{Ad}
\exp\left[\frac{t}{a^2}\left(\frac{1}{4}
-X^2\right)\right]
\tr_{T_1}\Psi\left(\frac{t}{a^2};2i(X-E)\right)\,.
\nonumber\\
\eea
Thus, the total Yang-Mills heat trace is
\bea
\Theta_{YM}(t)&=&
16\pi^3 a^2a_1a_2
\Omega\left(\frac{t}{a_1^2}\right)
\Omega\left(\frac{t}{a_2^2}\right)
W_{YM}\left(\frac{t}{a^2}\right)\,,
\eea
where
\bea
W_{YM}(t)
&=&
\int\limits_{C}
\frac{d\omega}{\sqrt{4\pi t}}\;
\exp\left\{-\frac{\omega^2}{4t}\right\}
\frac{\omega/2}
{\sin\left[\omega/2\right]}
\nonumber\\
&&
\times
\tr_{Ad}\Biggl\{
\exp\left[t\left(\frac{1}{4}-X^2\right)-X\omega\right]
\left[\tr_{T_1}\exp(E\omega)-2\right]
\Biggr\}\,.
\eea

The trace over the vector representation can be easily computed. By using
\be
\exp(E\omega)=\II-P+P\cos\omega+E\sin\omega\,,
\ee
we get
\be
\tr_{T_1}\exp(E\omega)=2+2\cos\omega\,.
\ee
Thus, the function $W_{YM}(t)$ is
\bea
W_{YM}(t)&=&
\int\limits_{C}
\frac{d\omega}{\sqrt{4\pi t}}\;
\exp\left\{-\frac{\omega^2}{4t}\right\}
\frac{\omega/2}
{\sin\left[\omega/2\right]}2\cos\omega
%\nonumber\\
%&&
%\times
\tr_{Ad}
\exp\left[t\left(\frac{1}{4}-X^2\right)-X\omega\right]
\,.
\nonumber\\
\eea

Recall that the eigenvalues of the matrix $X$ 
are imaginary half-integers
\be
{\rm Spec}\,(X)=\left\{
\underbrace{0,\dots,0}_{r},
i\frac{n_1}{2},-i\frac{n_1}{2},\dots, i\frac{n_p}{2}, -i\frac{n_p}{2}, 
\right\}\,,
\ee
where $p$ is the number of positive roots, $r$ is the rank of the group, 
and $n_1, \dots, n_p$ are some 
integers that we call monopole numbers.

Therefore,
the trace over the adjoint representation can be also computed to
take more explicit form
\bea
W_{YM}(t)&=&
\sum_{j=1}^p
\exp\left[\frac{t(1+n_j^2)}{4}\right]
\int\limits_{C}
\frac{d\omega}{\sqrt{4\pi t}}\;
\exp\left\{-\frac{\omega^2}{4t}\right\}
\frac{\omega/2}
{\sin\left[\omega/2\right]}
%\nonumber\\
%&&
%\times
4\cos\left[n_j\omega/2\right]\cos\omega
\nonumber
\\
&&
+r
\exp\left(\frac{t}{4}\right)
\int\limits_{C}
\frac{d\omega}{\sqrt{4\pi t}}\;
\exp\left\{-\frac{\omega^2}{4t}\right\}
\frac{\omega/2}
{\sin\left[\omega/2\right]}
2\cos\omega\,,
%\nonumber
%\\
\eea
which can be written in terms of the function $\Psi(t;n)$
\bea
W_{YM}(t)&=&
2\sum_{j=1}^p
\exp\left[\frac{t(1+n_j^2)}{4}\right]
\left\{
\Psi(t;2+|n_j|)+\Psi(t;2-|n_j|)
\right\}
%\nonumber\\
%&&
+2r
e^{t/4}\Psi(t;2)
\,.
\nonumber\\
\eea

The ultraviolet (high-energy) properties of the effective action depend of the
asymptotics of the function $W_{YM}$ as $t\to 0$.
By using the asymptotics of the function $\Psi(t;n)$ we obtain
\bea
W_{YM}(t)&\sim& 2m+C_1t+C_2t^2+\cdots
\,,
\eea
where
\bea
C_1&=& -\frac{19}{6}p -\frac{4}{3}r -\frac{4}{3}\sum_{j=1}^p n_j^2 \,,
\\
C_2&=& \frac{4}{15}p +\frac{19}{30}r +\frac{11}{6}\sum_{j=1}^p n_j^2\,.
\eea
By using this result, we obtain easily 
\be
\beta_{YM}=16\pi^3\frac{1}{xy}\left(C_2-za^2C_1
+z^2 a^4 m\right)\,.
\ee

Now, by subtracting enough terms of the Taylor expansion at $t=0$ we obtain
the renormalized heat trace
\bea
\Theta^{\rm ren}_{YM}(a^2t)&=&
16\pi^3 \frac{a^4}{xy}
\left\{
\Omega\left(x^2t\right)
\Omega\left(y^2t\right)
W_{YM}(t)
%\nonumber\\
%&&
%\hspace{-1cm}
-e^{-ta^2\lambda}R_{YM}(t;a^2\lambda)
\right\}\,,
\eea
where
\be
R_{YM}(t;a^2\lambda)
=2m+\left(C_1+2ma^2\lambda\right)t
+\left(C_2+C_1a^2\lambda+ma^4\lambda^2\right)t^2\,.
\ee

Then the one-loop Yang-Mills effective action is
\be
\Gamma^{}_{(1)\,YM}(x,y,\mu,a^2z)=
-\frac{\pi}{2}\frac{1}{xy}
\left\{
\left(C_2-za^2C_1
+z^2a^4 m\right)\log\frac{\mu^2}{\lambda}
+\Phi_{YM}\left(x,y;a^2\lambda,a^2z\right)
\right\}
\,,
\ee
where
\bea
\Phi_{YM}(x,y;a^2\lambda,a^2z)&=&
\int_0^\infty \frac{dt}{t^3}e^{-ta^2z}
\left\{
\Omega\left(x^2t\right)
\Omega\left(y^2t\right)
W_{YM}(t)
%\nonumber\\
%&&
%\hspace{-1cm}
-e^{-ta^2\lambda}R_{YM}(t;a^2\lambda)
\right\}\,.
\nonumber\\
\eea

The convergence properties of this integral as $t\to \infty$ (infrared region, low energies) crucially
depends on the asymptotics of the function $W_{YM}(t)$ as $t\to \infty$.
Let 
\be
\lambda_0(n)
=1-|n|+\left|1-\frac{|n|}{2}\right|
=
\left\{
\begin{array}{ll}
\displaystyle
2 & \mbox{ for } n=0,\\[10pt]
\displaystyle
\frac{1}{2} & \mbox{ for } |n|=1,\\[10pt]
\displaystyle
-\frac{|n|}{2} & \mbox{ for } |n|\ge 2\,.
\end{array}
\right.
\ee
Let
\be
\lambda_{\rm min}=\min_{1\le j\le p}\lambda_0(n_j)
\ee
and
\be
c=\sum_{1\le j\le p \atop \lambda_0(n_j)=\lambda_{\rm min}} \left(1+|2-|n_j|\,|\right)\,,
\ee
where the summation goes over all $j$ such that $\lambda_0(n_j)=\lambda_{\rm min}$.
Then as $t\to \infty$
\bea
W_{YM}(t)&\sim&
2tc e^{-t\lambda_{\rm min}}
+6rt e^{-2t}+\cdots
\,.
\label{622}
\eea

Thus we see that for $|n_j|\ge 2$ the lowest eigenvalue is negative, which leads to the exponential growth of the function $W_{YM}(t)$ as $t\to\infty$ and, as a result, to the infrared divergence of the integral for the zeta-function and the effective action and to the instability of the chromomagnetic vacuum. The only stable configurations are 
those in which all monopole numbers are equal to $0$ or $\pm 1$.
Recall that the flat space limit 
$a\to \infty$ is recovered as $X=-a^2H\to\infty$, keeping $H$ constant. That is, in the limit $a\to \infty$ the matrix $|X|\to \infty$ and, of course, 
the monopole numbers $|n_j|\to \infty$ 
(which can be interpreted physically as condensation of magnetic monopoles)
leading to the instability. This is nothing but the instability of the Savvidy chromomagnetic vacuum in 
the flat Euclidean space.

However, as we have shown above since the monopole numbers $n_j$ are not independent but are rather determined by the roots of the Lie algebra 
$\gfrak$ it is impossible to have all monopole numbers bounded between $-1$ and $1$. Therefore, the lowest eigenvalue is necessarily negative and
is equal to
\be
\lambda_{\rm min}=-\frac{1}{2}n_{\rm max}\,,
\ee
where
$
n_{\rm max}=\max_{1\le j\le p}|n_j|\,
$
is the largest monopole number. Since $n_{\rm max}\ge 2$,
then the constant $c$ is now
\be
c=(n_{\rm max}-1)q\,,
\ee
where $q$ is the multiplicity of the largest monopole number.

Let us consider the group $SU(N)$ for simplicity. Then the monopole numbers are $n_{ij}=k_i-k_j$, where $k_i$ are some integers whose sum is equal to zero.
For $N=2$ we have just two non-zero integers, $k_1$ and $k_2$, whose sum is equal to zero. Then there is one non-zero monopole number
\be
n_1=2k_1\,.
\ee
Therefore, 
\be
n_{\rm max}=2|k_1|\ge 2 \qquad\mbox{and}\qquad 
\lambda_{\rm min}=-|k_1|\le -1.
\ee

For $N=3$ there are three non-zero integers $k_1, k_2, k_3$ whose sum is equal to zero. Then there are three non-zero monopole numbers
\be
n_1=k_1-k_2\,,\qquad
n_2=2k_1+k_2\,,\qquad
n_3=k_1+2k_2\,.
\ee
Then, if $k_1=k_2$,
\be
n_{\rm max}=3|k_1|\ge 3\,.
\ee
If $k_1>k_2> 0$, then
\be
n_{\rm max}=2k_1+k_2\ge 3\,.
\ee
If $k_2< 0< 2|k_2|\le k_1$, then
\be
n_{\rm max}=2k_1-|k_2|\ge 3\,.
\ee
If $k_2<0<k_1$ and $|k_2|\le 2k_1\le 4|k_2|$, then
\be
n_{\rm max}=k_1+|k_2|\ge 3\,.
\ee
Finally, if 
If $k_2<0<k_1$ and $2k_1\le |k_2|$, then
\be
n_{\rm max}=2|k_2|-k_1\ge 3\,.
\ee
In this case 
\be
\lambda_{\rm min}\le -\frac{3}{2}\,.
\ee

More generally, one can show that for the group $SU(N)$ we have
$n_{\rm max}\ge 2$ for even $N$ and $n_{\rm max}\ge 3$ for odd $N$.
The equality is achieved by choosing the non-zero integers $k_1, \dots, k_N$, (whose sum is equal to zero) such that
there are as many as possible (monopole-antimonopole) 
pairs $(+1, -1)$. Then for even $N$, the largest difference $|k_i-k_j|$ is equal to $2$. For odd $N$, we will have as many pairs $(+1,-1)$ as possible and a triple $(2,-1,-1)$. Then the largest difference $|k_i-k_j|$ is equal to $3$.
Thus, the bottom eigenvalue for the group $SU(N)$ is less or equal to $(-1)$ 
for even $N$ and less or equal to $(-3/2)$ for odd $N$.

Let us compute this function for the group $SU(2N)$ so that 
$m=4N^2-1$, $r=2N-1$ and $p=N(2N-1)$, in the simple case when 
the numbers $k_j$ are chosen as $N$ pairs $(+1,-1)$, that is, 
$k_1=\cdots=k_{N}=1$ and $k_{N+1}=\cdots=k_{2N}=-1$. 
Then there are only $N^2$ nonzero monopole numbers which are equal to $2$ and the rest are equal to zero. Then
\bea
W^{SU(2N)}_{YM}(t)&=&
2N^2
e^{5t/4}
\left\{\Psi(t;4)+\Psi(t;0)
\right\}
%\\
%&&
+2(2N^2-1)
e^{t/4}\Psi(t;2)
%\nonumber
\\[5pt]
&=&
t\left[2N^2e^t+6N^2e^{-t}+6(2N^2-1)e^{-2t}\right]
\nonumber\\
&&
+\left[4N^2e^{5t/4}+2(2N^2-1)e^{t/4}\right]\Psi(t;4)
\,.
\label{740xx}
\nonumber
\eea
The asymptotics of this function as $t\to \infty$ are
\be
W^{SU(2N)}_{YM}(t)=2N^2te^{t}+O(te^{-t})\,.
\ee

A few remarks are in order.
First of all, the effective action does not depend on the parameter
$\lambda$; it can be chosen arbitrarily, for example, $\lambda=1/a^2$.
Second, for a sufficiently large ${\rm Re}\,z$ the function
$\Phi_{YM}(x,y;a^2\lambda,a^2z)$, as a function of $z$, 
is analytic. Since the function $W_{YM}(t)$ grows exponentially as $t\to \infty$, 
the function $\Phi_{YM}$ has a branching singularity at
\be
z_{0}=-\frac{\lambda_{\rm min}}{a^2}=\frac{1}{2 a^2}n_{\rm max}\,.
\ee

%=========================================
\subsection{Effective Action of Matter Fields}

For the scalar operator the potential term is $Q_0$, which is determined by the second derivative of the scalar potential $V(\phi)$, and the generator $G_0$ is
\be
G_0=-X_0\,,
\ee
where
$X_{0}$ is the generator of the spin group ${\rm Spin}(2)$ taking values in the representation $W_{0}$ of the gauge group. 
Therefore, the heat trace of the operator $L_0$ is
\bea
\Theta_0(t) 
&=&
16\pi^3 a^2a_1a_2
\Omega\left(\frac{t}{a_1^2}\right)\Omega\left(\frac{t}{a_2^2}\right)
\tr_{W_0}
\exp\left[-\frac{t}{a^2}\left(X_0^2+a^2Q_0-\frac{1}{4}\right)\right]
\Psi\left(\frac{t}{a^2};-2iX_0\right)\,.
\nonumber\\
\eea

For the spinor operator 
the generator $G_{\rm spin}$ is now
\be
G_{\rm spin}=\frac{1}{2}\gamma-X_{\rm spin}\,,
\ee
where
$\gamma=\gamma_3\gamma_4$, with $\gamma_a$ being the Dirac matrices, so that 
$\frac{1}{2}\gamma$ is the generator of the spinor representation; of course,
$\gamma^2=-\II$, and $X_{\rm spin}=X_{\rm spin}(\Sigma_{34})$ is the generator of the spin group ${\rm Spin}(2)$ taking values in the representation $W_{\rm spin}$ of the gauge group.
The endomorphism $Q_{\rm spin}$ 
is
\be
Q_{\rm spin}=
\frac{1}{2a^2}\II
+\frac{1}{ a^2}\gamma X_{\rm spin} 
+ M^2\,.
\ee
Then
\be
G_{\rm spin}^2+a^2Q_{\rm spin}=\frac{1}{4}\II
+X_{\rm spin}^2+a^2M^2\,.
\ee
Therefore, the heat trace for the operator $L_{\rm spin}$ is
\bea
\Theta_{\rm spin}(t) 
&=&
16\pi^3 a^2a_1a_2
\Omega\left(\frac{t}{a_1^2}\right)\Omega\left(\frac{t}{a_2^2}\right)
\tr_{W_{\rm spin}}\tr_{T_{\rm spin}}
\exp\left[-\frac{t}{a^2}\left(X_{\rm spin}^2+a^2M^2\right)\right]
\nonumber\\
&&
\times
\Psi\left(\frac{t}{a^2};i(2X_{\rm spin}-\gamma)\right)\,.
%\nonumber\\
\eea

Thus the total heat trace for the matter fields reads
\bea
\Theta_{\rm mat}(t)&=&
16\pi^3 a^2a_1a_2
\Omega\left(\frac{t}{a_1^2}\right)\Omega\left(\frac{t}{a_2^2}\right)
W_{\rm mat}\left(\frac{t}{a^2}\right)\,,
\eea
where
\bea
W_{\rm mat}\left(t\right)
&=&
\int\limits_{C}
\frac{d\omega}{\sqrt{4\pi t}}\;
\exp\left\{-\frac{\omega^2}{4t}\right\}
\frac{\omega/2}
{\sin\left[\omega/2\right]}
\nonumber\\
&&\times
\Biggl\{\tr_{W_0}
\exp\left[-t\left(X_0^2+a^2Q_0-\frac{1}{4}\right)-X_0\omega\right]
\nonumber\\
&&
-\tr_{W_{\rm spin}}\tr_{T_{\rm spin}}
\exp\left[-t\left(X_{\rm spin}^2+a^2M^2\right)
+\left(\frac{1}{2}\gamma-X_{\rm spin}\right)\omega\right]
\Biggr\}\,.
%\nonumber
\eea

To proceed further, we assume that the matrix $M$ does not transform with
respect to the spinor representation (it does not have spinor indices, but only the group indices). Then
by using
\be
\exp\left[\frac{1}{2}\gamma\omega\right]=\cos(\omega/2)\II+\gamma
\sin(\omega/2)\,,
\ee
the trace over spinor representation is easily computed
\be
\tr_{T_{\rm spin}}
\exp\left[\frac{1}{2}\gamma\omega\right]
=4\cos(\omega/2)
\,.
\ee
Thus, we get
\bea
W_{\rm mat}(t)&=&
\int\limits_{C}
\frac{d\omega}{\sqrt{4\pi t}}\;
\exp\left\{-\frac{\omega^2}{4t}\right\}
\frac{\omega/2}
{\sin\left[\omega/2\right]}
\nonumber\\
&&\times
\Biggl\{\tr_{W_0}
\exp\left[-t\left(X_0^2+a^2Q_0-\frac{1}{4}\right)-X_0\omega\right]
\nonumber\\
&&
-4\tr_{W_{\rm spin}}
\exp\left[-t\left(X_{\rm spin}^2+a^2M^2\right)
-X_{\rm spin}\omega\right]\cos(\omega/2)
\Biggr\}\,.
%\nonumber
\eea

The eigenvalues of the matrices $X_0$ and $X_{\rm spin}$ must be
imaginary half-integers
\be
{\rm Spec}(X_{0})=\left\{
i\frac{m_1}{2},\dots, i\frac{m_{N_{0}}}{2}, 
\right\}\,,
\ee
\be
{\rm Spec}(X_{\rm spin})=\left\{
i\frac{k_1}{2},\dots, i\frac{k_{N_{\rm spin}}}{2}, 
\right\}\,,
\ee
where $m_i$ and $k_j$ are some integers determined by the 
weights of the representations  realized
by spinors and scalars. 

For simplicity, we assume that the symmetry is not broken
so that all scalar fields have the same mass $m_0$ and all spinor fields have mass $m_{\rm spin}$. Then the potential terms $Q_0$
and $M$ are proportional to the identity, that is,
\be
Q_{0}=m_0^2\II\,,\qquad
M=m_{\rm spin}\II\,.
\ee
In this case
\bea
W_{\rm mat}(t)&=&
\exp\left[\left(\frac{1}{4}-a^2m_0^2\right) t\right]
\sum_{j=1}^{N_0}
e^{tm_j^2/4}
\Psi(t;m_j)
\nonumber\\
&&
-2e^{-ta^2m_{\rm spin}^2}\sum_{j=1}^{N_{\rm spin}}
e^{tk_j^2/4}
\left[\Psi(t;1-k_j)+\Psi(t;1+k_j)\right]
\,.
\eea

The leading asymptotics of this function as $t\to \infty$ is
\bea
W_{\rm mat}(t)\sim
r_0t e^{-ta^2m_0^2}
-2\sum_{j=1}^{N_{\rm spin}}|k_j|\;
te^{-ta^2m_{\rm spin}^2}
\,,
\eea
where $r_0$ is the number of zeros among the scalar weights $m_j$.

This indicates infrared 
instability for massless fields, when $m_0=m_{\rm spin}=0$.
It is interesting to note that
the infrared instability of massless spinor fields
is not caused by the zero numbers $k_j$; in fact, that contribution 
is proportional to $\Psi(t;1)$ and has a nice
exponentially decreasing behavior at $t\to \infty$. This instability is intrinsic and cannot be avoided if there are non-zero numbers $k_j$.
This simply means that massless Dirac operator with a constant non-zero magnetic field always has zero modes on $S^2$. We can even compute the multiplicity of the zero eigenvalue, that is, the dimension of the kernel of massless Dirac operator,
\be
{\rm dim}\, {\rm Ker}\, L_{\rm spin}\Bigg|_{m_{\rm spin}=0}
=2\sum_{j=1}^{N_{\rm spin}}|k_j|
=4\tr_{W_{\rm spin}}|X_{\rm spin}|\,.
\ee

Let us compute the contribution of spinors in the fundamental 
representation of the group $SU(2N)$
when the numbers $k_1, \dots, k_{2N}$ are chosen as
$k_1=\dots=k_N=1$ and $k_{N+1}=\dots=k_{2N}=-1$, so that $N_{\rm spin}=2N$. 
Then 
\bea
W^{SU(2N)}_{\rm spin}(t)&=&
-4N\exp\left[\left(\frac{1}{4}-a^2m_{\rm spin}^2\right)t\right]
\left[\Psi(t;2)+\Psi(t;0)\right]
\nonumber\\
&=&
-4Nte^{-a^2m_{\rm spin}^2t}
-8N\exp\left[\left(\frac{1}{4}-a^2m_{\rm spin}^2\right)t\right]
\Psi(t;2)
\,.
\label{768xx}
\eea

Now, the heat trace for matter fields can be renormalized in the same fashion as we have done for Yang-Mills fields. 
Then we get the effective action for matter fields exactly in the same form as for the Yang-Mills fields
with the function $\Phi_{\rm mat}(x,y,a^2\lambda,a^2z)$ that is expressed in terms of the function $W_{\rm mat}(t)$
exactly in the same way as for the Yang-Mills fields.

%=============================================
\section{Thermodynamics of Yang-Mills Theory}

In this section we investigate the entropy and the heat capacity of the
gluon gas. 
The temperature is related to the radius of the first circle
by
$
T=\frac{1}{2\pi a_1}\,.
$
The volume of the space is expressed in terms of the radius of the second circle and the radius of the sphere
$V=8\pi^2 a_2 a^2
\,.
$

For a canonical statistical ensemble with
fixed $T$ and $V$
the free energy $F=E-TS$ is 
a function of $T,V$ defined by
\be
F=T\Gamma=\frac{1}{2\pi}\frac{\Gamma}{a_1}\,.
\ee
Then the entropy is defined by
\be
S=-\frac{\partial }{\partial T}F
=a_1^2 \frac{\partial }{\partial a_1}
\frac{\Gamma}{a_1}
\ee
and the heat capacity at constant volume is
\bea
C_v&=&T\frac{\partial}{\partial T}S
%\nonumber\\
=
-a_1^2\left(a_1\frac{\partial^2 }{\partial a_1^2}
\frac{\Gamma}{a_1}
+
2\frac{\partial}{\partial a_1}
\frac{\Gamma}{a_1}
\right)
\,.
\eea

By using our results for the effective action we obtain
\bea
\frac{\Gamma}{a_1}&=&
16\pi^3\sigma
\frac{a_2}{a^2}
-\frac{\pi}{2}\frac{a_2}{a^2}
\left\{
b\log\frac{\mu^2}{\lambda}
+\Phi\left(\frac{a}{a_1},\frac{a}{a_2};a^2\lambda,a^2z\right)
\right\}
\,,
\eea
where the function $\Phi$ is given by the sum of the corresponding functions for the
Yang-Mills fields and the matter fields,
$\Phi=\Phi_{YM}+\Phi_{\rm mat}$, and
\bea
b&=&\frac{1}{16\pi^3}
\frac{a^2}{a_1a_2}
\left(\beta_{YM}+\beta_{\rm mat}\right)
\,.
\eea

It is easy to see that 
neither the classical part nor the coefficient $b$ depend on the temperature.
Therefore,
the entropy and the heat capacity 
do not depend neither on the classical term nor on the renormalization 
parameter $\mu$. Recall also that the parameter $\lambda$ is completely arbitrary;
so we can set it equal to $\lambda=1/a^2$.
Thus the entropy and the heat capacity of the quark-gluon gas are given by the derivatives
of the functions $\Phi$
\bea
S&=&\frac{\pi }{2y}
\Phi_x(x,y,1,a^2z)\,,
\\
C_v&=&\frac{\pi }{2}\frac{x}{y}
\Phi_{xx}(x,y,1,a^2z)
=\frac{\pi }{2}\frac{a_2}{a_1}
\Phi_{xx}(x,y,1,a^2z)\,.
\eea
The entropy and the heat capacity per unit volume are
\bea
\frac{S}{V}&=&\frac{1}{16\pi a^3}
\Phi_x(x,y,1,a^2z)\,,
\\
\frac{C_v}{V}&=&\frac{1}{16\pi a_1 a^2}
\Phi_{xx}(x,y,1,a^2z)\,.
\eea

Now, by differentiating the function 
$\Phi$ with respect to $x$ we get
\bea
\Phi_x(x,y;1,a^2z)&=&2x
\int_0^\infty \frac{dt}{t^2}e^{-ta^2z}
\Omega'\left(x^2t\right)
\Omega\left(y^2t\right)
W(t)\,,
\label{878xx}
\\
\Phi_{xx}(x,y;1,a^2z)&=&
2\int_0^\infty \frac{dt}{t^2}e^{-ta^2z}
\left\{
\Omega'\left(x^2t\right)
+2x^2 t\Omega''\left(x^2t\right)
\right\}
\Omega\left(y^2t\right)
W(t)
\nonumber\\
&=&
2x^2\int_0^\infty \frac{dt}{t^2}
e^{-ta^2z/x^2}
\left\{
\Omega'\left(t\right)
+2t\Omega''\left(t\right)
\right\}
\Omega\left(\frac{y^2}{x^2}t\right)
W\left(\frac{t}{x^2}\right)
\,,
\nonumber\\
\label{881xx}
\eea
where
\be
W(t)=W_{YM}(t)+W_{\rm mat}(t)\,.
\ee

Now by using the asymptotics of the function $\Omega$
we easily see that the integrals (\ref{878xx}) and (\ref{881xx})
converge at $t\to 0$. Moreover,
for large $z$ these integrals also converge 
as $t\to \infty$. 
Since the function $W_{YM}$ increases
exponentially at infinity, the function $\Phi$ has a singularity
at a finite positive value of $z$. However, because of the asymptotics
of the function $\Omega'(t)+2t\Omega''(t)$ as $t\to \infty$,
(\ref{888xx}), we immediately see that the integral (\ref{881xx}) for the heat capacity
may converge even in the infrared limit $z\to 0$, due to the presence
of an extra exponential factor $e^{-t}$. Namely, for the groups $SU(2N)$
there are monopole configurations such that the function $W_{YM}(t/x^2)$
increases as $e^{t/x^2}$, and therefore, the integral would converge
for sufficiently large $x$.

That is why we investigate this case in more detail. We set $z=0$ 
(and $y=0$, for simplicity) to get
\bea
\Phi_{xx}(x,0;1,0)&=&
2x^2\int_0^\infty \frac{dt}{t^2}
\left\{
\Omega'\left(t\right)
+2t\Omega''\left(t\right)
\right\}
W\left(\frac{t}{x^2}\right)
\,.
\label{891xx}
\eea

We consider the group $SU(2N)$ when the function $W$
is given by (\ref{740xx}) and (\ref{768xx}) for the spinors.
We decompose it according to
\be
W\left(\frac{t}{x^2}\right)=2N^2\frac{t}{x^2}e^{t/x^2}
-4Ns\frac{t}{x^2}
+V\left(\frac{t}{x^2}\right)\,,
\label{892xx}
\ee
where $s=0$ for massive spinors and $s=1$ for massless spinors,
when $m_{\rm spin}=0$, and
the function $V$ is exponentially decreasing as $t\to \infty$.

We now split the integral (\ref{891xx}) into three parts
accordingly 
\be
\Phi_{xx}(x,0;1,0)=I_1+I_2-8Ns\nu_1\,,
\ee
where
\be
\nu_1=\int_0^{\infty}\frac{dt}{t}
\left\{
\Omega'\left(t\right)
+2t\Omega''\left(t\right)
\right\}\,,
\ee
and
\bea
I_1&=&
4N^2\int_0^{\infty}\frac{dt}{t}e^{t/x^2}
\left\{
\Omega'\left(t\right)
+2t\Omega''\left(t\right)
\right\}\,,
\\
I_2&=&
2x^2\int_{0}^{\infty}\frac{dt}{t^2}
\left\{\Omega'\left(t\right)
+2t\Omega''\left(t\right)
\right\}V\left(\frac{t}{x^2}\right)
\,.
\eea
Recall that $x=a/a_1=2\pi a T$; so $x\to \infty$ is the high-temperature limit and $x\to 0$ is the limit of zero temperature.

We consider the high-temperature limit first. As $x\to \infty$ the second integral is
\be
I_2\sim 4(4N^2-1)\nu_2 x^2\,,
\ee
where
\be
\nu_2=\int_0^{\infty}\frac{dt}{t^2}
\left\{
\Omega'\left(t\right)
+2t\Omega''\left(t\right)
\right\}\,.
\ee
For the first integral
in the limit $x\to \infty$ we get
\be
I_1\sim 4N^2\nu_1\,.
\ee
Thus, as $T\to\infty$ we obtain
\be
\frac{C_v}{V}\sim 2\pi^2(4N^2-1)\nu_2 T^3+\cdots\,,
\ee 
which is similar to the black body photon gas.

The 
first integral exhibits much more interesting behavior
due to the presence of the
growing exponential factor $e^{t/x^2}$. It converges at $t\to 0$ for any $x$. However, at $t\to\infty$ its convergence depends on the critical temperature. Recall that the other part of the 
integrand has an exponential factor $e^{-t}$. Therefore, the integral converges in the infrared domain $t\to \infty$ if
$x>1$ and diverges otherwise. This defines the {\it critical temperature}
\be
T_c=\frac{1}{2\pi a}\,.
\ee
At the critical temperature $T\approx T_c$ the second integral
is simply
\be
I_2=
2\int_{0}^{\infty}\frac{dt}{t^2}
\left\{\Omega'\left(t\right)
+2t\Omega''\left(t\right)
\right\}V\left(t\right)\,.
\ee

However, the first integral is nonanalytic near $T_c$, namely,
by using the asymptotics (\ref{888xx}), we get as $x\to 1^+$
\bea
I_1\sim 8N^2\left(1-\frac{1}{x^2}\right)^{-3/2}
\sim \frac{4N^2}{\sqrt{2}}
\left(x-1\right)^{-3/2}
\,.
\eea
Therefore, the heat capacity near the critical temperature, $T\to T_c^+$, is
\be
\frac{C_v}{V}\sim \frac{N^2}{4\pi \sqrt{2} a^3}
\left(\frac{T-T_c}{T_c}\right)^{-3/2}
\,.
\ee
This indicates the second-order
phase transition
with the critical exponent $\alpha=3/2$.

%==============================================================
%==============================================================
\section{Conclusion}

The primary goal of this paper was to study of the Yang-Mills vacuum in the low-energy 
(long-distance) limit. The Savvidy model of such a vacuum 
with constant chromomagnetic field in Minkowski spacetime suffers from a well-known instability, which exhibits itself in negative eigenvalues of the gluon operator. We noticed that a positive spatial curvature of the spacetime manifold 
acts as an effective mass term and, therefore, can stabilize the Savvidy vacuum.
That is why we considered the case of a compact spacetime manifold of the form
$S^1\times S^1\times S^2$ with a covariantly constant chromomagnetic Yang-Mills field on the sphere
$S^2$. On the sphere, such a configuration is of monopole type parametrized by a collection of half-integers (monopole numbers). Such a configuration has a well-defined (Savvidy type) flat space limit with constant chromomagnetic field when the radius $a$ of the 2-sphere $S^2$, as well as the monopole numbers, $n_j$, go to infinity, $n_j, a\to\infty$, such that the ratio $n_j/a^2$ (which defines the magnetic field) remains constant. This limit can be interpreted physically as the condensation of monopoles.

We computed exactly the spectra and the trace of the heat kernels of all relevant
operators, which enabled us to compute exactly the one-loop effective action.
We have found that the gluon operator does not have negative eigenvalues only when the
monopole numbers are between $-1$ and $1$. That is, any monopole number $n_j$, 
with $|n_j|\ge 2$, leads to an instability of the chromomagnetic vacuum. This confirms once again that the flat space limit with constant chromomagnetic field is unstable since it is created by infinitely large monopole numbers, formally $|n_j|\to \infty$. We showed that
for any compact simple gauge group there are always monopole numbers
with absolute value greater or equal to $2$, which means that {\it there is
no stable constant chromomagnetic configuration also in curved space}
(at least in the model $S^1\times S^1\times S^2$).

We also studied the thermal properties of Yang-Mills theory, in particular, we computed
the entropy and the heat capacity of the quark-gluon gas.
We have found that the heat capacity is well defined even in the infrared limit and
computed the high-temperature asymptotics of the heat capacity. Moreover, in a particular model
$SU(2N)$ 
we found
that the heat capacity has  a typical branching singularity $\sim (T-T_c)^{-3/2}$ 
at a finite critical temperature
$T_c=1/(2\pi a)$ indicating the second-order phase transition.

We conclude that to stabilize the chromomagnetic vacuum at lower energies one should consider
non-constant magnetic fields on non-compact spaces. Constant magnetic fields on compact symmetric
spaces are too rigid, they are completely determined by the spin connection and are of the same
order as the space curvature. This makes it impossible that the gluon operator with the potential term
$R^a{}_b-2{\cal F}^a{}_b$ is strictly positive. What one needs is a large Ricci tensor and a small
independent magnetic field to make this work.

It is also interesting to study the Yang-Mills vacuum on the 
Einstein model $S^1\times S^3$.
We intend to carry out such a study in future work.
Another interesting question to pose is whether the Yang-Mills vacuum
is stable when gravity is treated as a dynamical field. The technical calculations are not 
that difficult, but then, of course, we immediately face the 
non-renormalizability of general relativity, so we have to consider higher-derivative gravity instead. 

It is hard to imagine that this model can be directly 
relevant in hadron physics in the study of the confinement because of the completely
different energy scales dictated by the gravitational constant and the cosmological constant.
However, it can be relevant in the study of the structure of the
quark-gluon plasma in the early Universe.

%======================================================


\begin{thebibliography}{99}

\bibitem{avramidi91} 
I. G. Avramidi,  {\it A covariant technique for the calculation of the
one-loop effective  action}, Nucl. Phys. B {\bf 355} (1991) 712--754;
Erratum:  Nucl. Phys. B {\bf 509} (1998) 557-558. 

\bibitem{avramidi93}
I. G. Avramidi, {\it  A new algebraic approach for calculating the heat
kernel in gauge theories}, Phys. Lett. B {\bf 305} (1993) 27--34. 

\bibitem{avramidi95a} 
I. G. Avramidi,  {\it Covariant algebraic calculation of the one-loop effective
potential in non-Abelian gauge theory and a new approach to stability
problem},  J. Math. Phys. {\bf 36} (1995) 1557--1571.

\bibitem{avramidi99}
I. G. Avramidi, {\it A model of stable chromomagnetic vacuum in
higher-dimensional Yang-Mills theory},  Fortschr. Phys.,  {\bf 47}
(1999) 433--455.

\bibitem{avramidi00}
I. G. Avramidi, {\it Heat Kernel and Quantum Gravity},
(Berlin:  Springer,  2000)

\bibitem{avramidi08}
I. G. Avramidi,
{\it Heat kernel on homogeneous bundles}, 
Int. J. Geom. Meth. Mod. Phys., {\bf 5} (2008) 1-23

\bibitem{avramidi09}
I. G. Avramidi,
{\it Heat kernel on homogeneous bundles over symmetric spaces}, 
Commun. Math. Phys., {\bf 288} (2009) 963-1006.
%; DOI: 10.1007/s00220-008-0639-6 

\bibitem{avramidi10a}
I. G. Avramidi, 
{\it Non-perturbative effective action in gauge theories and quantum gravity},
Adv. Theor. Math. Phys. {\bf 14} (2010) 1--25.

\bibitem{avramidi10b}
I. G. Avramidi, 
{\it Mathemathical tools for calculation of the effective action in quantum gravity},
in: New Paths Toward Quantum Gravity, Ed. B. Booss-Bavnbek, G. Esposito and M. Lesch, (Berlin: Springer, 2010); pp. 193-259

\bibitem{dewitt65} 
B. S. De Witt, {\it Dynamical Theory of Groups and Fields}, (New York: 
Gordon and Breach,  1965).

\bibitem{frankel97} T. Frankel, {\it The Geometry of Physics}, (Cambridge: 
Cambridge University Press, 1997)


\bibitem{gilmore74} R. Gilmore, {\it Lie Groups, Lie Algebras and
Some of Their Applications}, (New York: Wiley, 1974)

\bibitem{nielsen78} 
N. K. Nielsen and P. Olesen, {\it  An unstable Yang-Mills mode}, Nucl.
Phys. {\bf B 144} (1978) 376--396.
 
\bibitem{nielsen79} 
H. B. Nielsen and P. Olesen, {\it A quantum liquid model for the QCD
vacuum: gauge and rotational invariance of domained and quantized
homogeneous color fields}, Nucl. Phys. {\bf B160} (1979) 380--396.

\bibitem{nikiforov88}
A. F. Nikiforov and V. B. Uvarov, {\it Special Functions of Mathematical Physics},
(Basel: Birkh\"auser, 1988)

\bibitem{savvidy77} 
G. K. Savvidy, {\it Infrared instability of the
vacuum state of gauge theories and asymptotic freedom}, Phys. Lett. 
{\bf B 71} (1977) 133--134.

\end{thebibliography}
\end{document}